\documentclass[twocolumn, tighten]{aastex62}

\usepackage[single=true]{acro}
\usepackage{ulem}
\usepackage{amsmath}	

\newcommand\taut{\ensuremath{\tau_{\mbox{\tiny{T}}}}}
\newcommand\sigmat{\ensuremath{\sigma_{\mbox{\tiny{T}}}}}
\newcommand\mbh{\ensuremath{M_{\mbox{\tiny{BH}}}}}

\newcommand\vff{{\ensuremath{v_{\mbox{\tiny{ff}}}}}}

\newcommand\vA{{\ensuremath{v_{\mbox{\tiny{A}}}}}}
\newcommand\vsh{{\ensuremath{v_{\mbox{\tiny{sh}}}}}}
\newcommand\Psh{\ensuremath{P_{\mbox{\tiny{sh}}}}}
\newcommand\ic{\ensuremath{{}_{\mbox{\tiny{IC}}}}}
\newcommand\ec{\ensuremath{\mbox{\tiny{EC}}}}
\newcommand\tildeb{\tilde{b}}
\newcommand\KN{\ensuremath{{}_{\mbox{\tiny{KN}}}}}
\newcommand\EBL{\ensuremath{{}_{\mbox{\tiny{EBL}}}}}
\newcommand\CR{\ensuremath{{}_{\mbox{\tiny{CR}}}}}
\newcommand\DSA{\ensuremath{{}_{\mbox{\tiny{DSA}}}}}
\newcommand\lar{{\ensuremath{\mbox{\tiny{L}}}}}
\newcommand\STO{\ensuremath{{}_{\mbox{\tiny{StA}}}}}

\DeclareAcronym{agn}{
  short = AGN,
  long  = active galactic nucleus,
  short-plural = s,
  long-plural-form = active galactic nuclei,
  class = astro,
  first-style = default
}
\DeclareAcronym{ebl}{
  short = EBL ,
  long  = extragalactic background light ,
  class = astro,
  first-style = default
}
\DeclareAcronym{smbh}{
  short = SMBH ,
  long  = supermassive black hole ,
  class = astro ,
  first-style = default
}
\DeclareAcronym{dsa}{
  short = DSA ,
  long  = diffusive shock acceleration ,
  class = astro ,
  first-style = default
}
\DeclareAcronym{sto}{
  short = StA ,
  long  = stochastic acceleration ,
  class = astro ,
  first-style = default
}
\DeclareAcronym{ic}{
  short = IC ,
  long  = inverse Compton ,
  class = astro ,
  first-style = default
}
\DeclareAcronym{kn}{
  short = KN ,
  long  = Klein--Nishina ,
  class = astro ,
  first-style = default
}
\DeclareAcronym{ec}{
  short = EC ,
  long  = elastic Coulomb ,
  class = astro ,
  first-style = default
}
\DeclareAcronym{pp}{
  short = \ensuremath{pp} ,
  long  = hadronuclear ,
  class = astro ,
  first-style = default
}
\DeclareAcronym{pg}{
  short = \ensuremath{p\gamma} ,
  long  = photomeson ,
  class = astro ,
  first-style = default
}
\DeclareAcronym{sed}{
  short = SED ,
  long  = spectral energy distribution ,
  class = astro ,
  first-style = default
}

\DeclareAcronym{alma}{
  short = ALMA ,
  long  = Atacama Large Millimeter/submillimeter Array ,
  class = instruments ,
  first-style = default
}

\graphicspath{{./}{figures/}}

\submitjournal{ApJ}

\shorttitle{On the high energy particles in massive black hole coronae}
\shortauthors{Inoue et al.}

\begin{document}

\title{On high-energy particles in accretion disk coronae of supermassive black holes:\\
implications for MeV gamma rays and high-energy neutrinos from AGN cores}

\author[0000-0002-7272-1136]{Yoshiyuki Inoue}
\email{yoshiyuki.inoue@riken.jp}
\affiliation{Interdisciplinary Theoretical \& Mathematical Science Program (iTHEMS), RIKEN, 2-1 Hirosawa, Saitama 351-0198, Japan}
\affiliation{Kavli Institute for the Physics and Mathematics of the Universe (WPI), UTIAS, The University of Tokyo, Kashiwa, Chiba 277-8583, Japan}

\author{Dmitry Khangulyan}%
\affiliation{Department of Physics, Rikkyo University, Nishi-Ikebukuro 3-34-1, Toshima-ku, Tokyo 171-8501, Japan}%

\author{Susumu Inoue}
\affiliation{Interdisciplinary Theoretical \& Mathematical Science Program (iTHEMS), RIKEN, 2-1 Hirosawa, Saitama 351-0198, Japan}

\author{Akihiro Doi}
\affiliation{Institute of Space and Astronautical Science JAXA, 3-1-1 Yoshinodai, Chuo-ku, Sagamihara, Kanagawa 252-5210, Japan}%
\affiliation{Department of Space and Astronautical Science, The Graduate University for Advanced Studies (SOKENDAI),3-1-1 Yoshinodai, Chuou-ku, Sagamihara, Kanagawa 252-5210, Japan}%

\begin{abstract}
Recent observations with ALMA have revealed evidence for non-thermal synchrotron emission from the core regions of two nearby Seyfert galaxies. This suggests that the coronae of accretion disks in \acp{agn} can be conducive to the acceleration of non-thermal electrons, in addition to the hot, thermal electrons responsible for their X-ray emission through thermal Comptonization. Here we investigate the mechanism of such particle acceleration, based on observationally inferred parameters for \ac{agn} disk coronae. One possibility to account for the observed non-thermal electrons is diffusive shock acceleration, as long as the gyrofactor $\eta_g$ does not exceed $\sim10^6$. These non-thermal electrons can generate gamma rays via inverse Compton scattering of disk photons, which can appear in the MeV band, while those with energies above $\sim100$~MeV would be attenuated via internal $\gamma\gamma$ pair production. The integrated emission from all \acp{agn} with thermal and non-thermal Comptonization can reproduce the observed cosmic background radiation in X-rays as well as gamma-rays up to $\sim 10$~MeV. Furthermore, if protons are accelerated in the same conditions as electrons and $\eta_g\sim30$, our observationally motivated model is also able to account for the diffuse neutrino flux at energies below 100--300 TeV. The next generation of MeV gamma-ray and neutrino facilities can test these expectations by searching for signals from bright, nearby Seyfert galaxies such as NGC~4151 and IC~4329A.
\acresetall
\end{abstract}


\section{Introduction} \label{sec:intro}
\Acp{agn} are powered by mass accretion onto \acp{smbh}. They emit intense electromagnetic radiation in broad range of frequencies. Measurements of X-ray spectra of \acp{agn} allow us to study various aspect of \acp{smbh} such as black hole spins \citep[e.g.,][]{Reynolds2014}, geometrical structures \citep[e.g.,][]{Ramos_Almeida2017}, and cosmological evolution \citep[e.g.,][]{Ueda2014}. 

A key for understanding these phenomena is primary X-ray radiation of the accretion disk which arises from Comptonization of disk photons in moderately thick thermal plasma, namely coronae, above an accretion disk \citep[see, e.g.,][]{Katz1976,1977A&A....59..111B,Pozdniakov1977,Galeev1979,Takahara1979,Sunyaev1980}. X-ray observations have indicated the coronal temperature of $\sim10^9$~K and the Thomson scattering opacity of $\gtrsim1$ \citep[e.g.][]{Zdziarski1994,Fabian2015}. However, the nature of \ac{agn} coronae is still veiled in mystery.
 
 Very recently, \citet{Inoue2018} has reported the detection of coronal radio synchrotron emission from two nearby Seyferts \citep[e.g.,][]{DiMatteo1997,Inoue2014,Raginski2006} utilizing the \ac{alma}. The inferred coronal magnetic field strength was $\sim10$~G with a size of $40R_s$, where $R_s$  is the Schwartzschild radius, for both active \acp{smbh} with a mass of $\sim10^8M_\odot$. It is also found that coronae of Seyferts contain both thermal and non-thermal electrons. This implies that acceleration of high energy particles happens in AGN coronae. 
 
 High energy particles in the nuclei of Seyferts have been discussed for a long time\footnote{High energy particles in the coronae of X-ray binaries have been also discussed in literature \citep[e.g.,][]{Bhattacharyya2003,Bhattacharyya2006}.}. In the past, it was argued that primary X-ray emission comes from pair cascades induced by high energy particles accelerated in and/or around accretion flows  \citep[e.g.,][]{Zdziarski1986,Kazanas1986, Ghisellini2004}. In the pair cascade model, particles are accelerated by shock dissipation in accretion flows \citep[e.g.,][]{Cowsik1982,Protheroe1983,Zdziarski1986,Kazanas1986,Sikora1987,Begelman1990}. However, the detection of the \ac{agn} spectral cutoffs \citep[e.g.,][]{Madejski1995,Zdziarski2000} and non-detection of Seyfert \acp{agn} in the gamma-ray band \citep[e.g.,][]{Lin1993} ruled out the pair cascade scenario as a dominant source for the primary X-ray emission\footnote{TeV gamma rays are measured from the Galactic center \citep{HESS2016}. This detection indicated possible particle acceleration in accretion flow, even though accretion rate in the Galactic center is several orders of magnitude lower than that in standard disks.}. 

In this paper, we investigate the production mechanism of the observed high energy particles in \ac{agn} coronae. As an example, we consider those high energy particles are supplied by \ac{dsa} processes \citep[e.g.,][]{Drury1983,Blandford1987} in the coronae. Contrary to the previously discussed AGN accretion shock models, the required shock power is much lower in order to explain the observed non-thermal species and to be in concordance with the current picture of coronal X-ray emission. Moreover, previous studies of high energy particles in \ac{agn} accretion disks have treated as free parameters corona size and magnetic field, which are important parameters for the understandings of particle acceleration. The \ac{alma} observations allowed us to determine both of them \citep{Inoue2018}. Most critically, the observationally determined strength of the magnetic field appeared to be significantly smaller than the one previously considered in the literature.  We take into account these newly determined coronal parameters.

Thermal coronal emission from Seyferts is known to explain the entire cosmic X-ray background radiation \citep[e.g.,][]{Ueda2014}. In contrast, the origin of the cosmic MeV background radiation from 0.1~MeV to several tens MeV is still unknown \citep[see e.g.,][]{Inoue2014_Fermi}. Here, the non-thermal electrons in coronae seen by ALMA will invoke power-law MeV gamma-ray emission via Comptonization of disk photons. Such non-thermal emission is suggested as a possible explanation for the cosmic MeV gamma-ray background radiation \citep{Inoue2008}. However, non-thermal electron species in the previous work were included in an ad hoc way. In this work, we revisit the contribution of Seyferts to the MeV gamma-ray background radiation by considering the particle acceleration of non-thermal populations in coronae together with the latest X-ray luminosity function of Seyferts \citep{Ueda2014}. 

High energy particles around accretion disks of \acp{agn} also generate intense neutrino emission through \ac{pp} and \ac{pg} interaction processes by interacting accreting gas and photon fields \citep[e.g.,][]{Eichler1979,Begelman1990,Stecker1992,Alvarez-Muniz2004}. Although these originally predicted fluxes have been significantly constrained by high energy neutrino observations \citep{IceCube2005}, recent studies have revisited the estimated fluxes and found that \ac{agn} core models are still viable \citep{Stecker2005,Stecker2013,Kalashev2015}. However, normalization of neutrino fluxes from \acp{agn} and acceleration properties of high energy particles in those models are assumed to match with the observation. In this work, we also discuss the possible contribution from \ac{agn} cores given our \ac{alma} observations and investigate the required parameter spaces for the explanation of the IceCube diffuse neutrino fluxes.

We describe general particle acceleration processes in \ac{agn} coronae in \S~\ref{sec:acceleration}. The broadband emission spectrum of the central region of \acp{agn} and physical properties of \ac{agn} coronae are presented in \S~\ref{sec:property}. Relevant timescales and steady-state particle spectra are discussed in \S~\ref{sec:timescales} and \S~\ref{sec:particle_spectrum}, respectively. \S~\ref{sec:g_nu_AGN} and \S~\ref{sec:background} present the results of the expected gamma-ray and neutrino fluxes from individual AGN cores and the cosmic gamma-ray and neutrino background fluxes from \ac{agn} cores, respectively.  Discussion including other possible particle acceleration mechanism is given in \S~\ref{sec:discussion}, and conclusions are in \S~\ref{sec:conclusion}. Throughout this paper, we adopt the standard cosmological parameters of $(h, \Omega_M, \Omega_\Lambda) = (0.7, 0.3, 0.7)$.

\section{Particle Acceleration in Nuclei of Seyferts}
\label{sec:acceleration}
As non-thermal coronal synchrotron emission is seen in nearby Seyferts \citep{Inoue2018}, particle acceleration should occur in AGN coronae, even though thermal populations are energetically dominant. Particle acceleration mechanism in the coronae is highly uncertain. Various acceleration mechanisms can take place in the coronae such as  \ac{dsa} mechanism \cite[e.g.,][]{Drury1983,Blandford1987}, turbulent acceleration \citep[e.g.,][]{Zhdankin2018}, magnetosphere acceleration \citep[e.g.,][]{Beskin1992,Levinson2000}, and magnetic reconnection \citep[e.g.,][]{Hoshino2012}. In this work, for simplicity, we consider the \ac{dsa} as the fiducial particle acceleration process. We discuss the other possible acceleration processes in \S~\ref{sec:other_acc}. 

In order to investigate particle acceleration mechanism of the observed non-thermal electrons, we consider the interaction of locally injected relativistic particles with the matter, photons, and magnetic field in the infalling coronae. Although the location of shock sites is uncertain, for simplicity, we assume that shocks occur inside of the coronae. The shock accelerates a part of inflow plasma to high energies. As the energy loss timescale of high energy protons is in general longer than the free-fall timescale, a sufficiently high energy density of relativistic particles is maintained to provide pressure to support a standing shock around a \ac{smbh} \citep{Protheroe1983}. 

Coronae are assumed to be spherical with a radius of $R_c\equiv r_cR_s$. $r_c$ is the dimensionless parameter of the corona size and $R_s=2G\mbh/c^2$, where $G$ is the gravitational constant, $\mbh$ is the mass of the central \ac{smbh}, $c$ is the speed of light. Coronae are also set to be in a steady state. We also do not consider positrons in coronae. Thus, the proton number density $n_p$ is equal to the electron density $n_e$ in this work, which gives the maximum number of protons in coronae. $n_e$ is defined through the Thomson scattering opacity in coronae, \(\taut\)  as 
\begin{eqnarray}
	n_e &=& \frac{\taut}{\sigmat R_c}\\ \nonumber
	&\simeq& 1.4\times10^9\left(\frac{\taut}{1.1}\right)\left(\frac{r_c}{40}\right)^{-1}\left(\frac{\mbh}{10^8M_\odot}\right)^{-1}\ {\rm cm}^{-3},
\end{eqnarray}
 where $\sigmat$ is the Thomson scattering cross section.

\subsection{Dynamical Timescale}
The gas is assumed to be spherically accreted on to the \ac{smbh} with free-fall velocity $\vff = \sqrt{2G\mbh/R_c}$. The free-fall timescale from the coronal region is estimated to be
\begin{equation}
\label{eq:t_fall}
t_{\rm fall}= R_c / \vff\simeq2.5\times10^5\left(\frac{r_c}{40}\right)^{1/2}\left(\frac{\mbh}{10^8M_\odot}\right)\ [{\rm s}].
\end{equation}

\subsection{Radiative Cooling}
High energy particles loose their energies through radiative cooling processes. In \ac{agn} coronae, high-energy electrons mainly lose their energies via synchrotron and \ac{ic} radiation. The synchrotron cooling rate for an electron with a Lorentz factor of $\gamma_e$ is
\begin{eqnarray}
 t_{{\rm syn}, e}(\gamma_e) &=& \frac{3}{4} \frac{m_e c}{\sigmat  U_{\rm B}} \gamma_e^{-1}, \\ \nonumber
 &\simeq& 7.7\times10^4\left(\frac{B}{10~{\rm G}}\right)^{-2}\left(\frac{\gamma_e}{100}\right)^{-1}\ [{\rm s}],
\end{eqnarray}
where $m_e$ is the electron rest mass and $U_{\rm B} =B^2/8\pi$ is the magnetic field energy density of magnetic field strength $B$.

The inverse Compton cooling rate including the \ac{kn} cross section \citep{Jones1968,Moderski2005,Khangulyan2014} is
\begin{equation}
\label{eq:time_ic}
t\ic(\gamma_e) = \frac{3 m_e c}{4\sigmat  }\left[\int\limits_0^{\infty}d\epsilon f\KN(\tildeb)\frac{U_{\rm ph}(\epsilon)}{\epsilon} \right]^{-1}\gamma_e^{-1},
\end{equation}
where $\tildeb\equiv 4\gamma_e\epsilon/m_ec^2$ and $f\KN \simeq 1/(1.0+\tildeb)$ \citep{Moderski2005}. $\epsilon$ is the target photon energy and $U_{\rm ph}$ is the photon energy density given as $U_{\rm ph}(\epsilon)=L_{\rm ph}(\epsilon)/4\pi R_c^2c$.  The total \ac{agn} disk luminosity, $L_{\rm ph}$, which includes contribution from the accretion disk and corona, is defined in \S~\ref{sec:AGN_SED}. For simplicity, we consider a uniform photon density in the coronae. If the coronae has spatially homogeneous emissivity rather uniform emission, the mean photon density inside the source is enhanced by a factor of $\sim2.24$ on average \citep{Atoyan1996}. 

For the typical characteristics of the coronae, the energy density of the photon field is
\begin{eqnarray}
  &&U_{\rm ph,}{}_{\rm tot}=\int d \epsilon\, U_{\rm ph}(\epsilon)\\
  && \sim5\times10^3 \frac{L_{\rm ph,bol}}{2\times10^{45}\rm\,erg\,s^{-1}} \left(\frac{r_c}{40}\right)^{-2}\left(\frac{\mbh}{10^8M_\odot}\right)^{-2}[{\rm erg\,cm^{-3}}]\,.\nonumber
\end{eqnarray}
For the magnetic field strength inferred with \ac{alma}, \(B\simeq10\rm\,G\) for $\mbh=10^8M_\odot$ SMBHs, the energy density of the photon field exceeds the magnetic field energy density if \mbox{\(L_{\rm ph,bol}\geq2\times10^{42}\rm\,erg\,s^{-1}\)}. We note that the dominance of photon fields over magnetic field does not necessary prevents particle acceleration as such conditions are met in some efficient non-thermal sources, e.g., in gamma-ray binary systems \citep{2006JPhCS..39..408A,2008MNRAS.383..467K}. Moreover, high density of target photons can enable the converter acceleration mechanism if a relativistic velocity jump present in the system \citep{2003PhRvD..68d3003D}. 

Relativistic protons are predominately cooled though inelastic \ac{pp} interactions, \ac{pg} reactions, and proton \ac{ic}/synchrotron channels. Since only the Thomson regime might be relevant for the proton \ac{ic} cooling,  the proton synchrotron and \ac{ic}  cooling time-scales are 
\begin{equation}
 t\ic{}_{{\rm /syn}, p} = \frac{3}{4} \left(\frac{m_p}{m_e}\right)^3\frac{m_e c^2}{c\sigmat  U_{\rm ph/B}} \gamma_p^{-1}\,,
 \end{equation}
where $m_p$ is the proton rest mass and $\gamma_p$ is the proton Lorentz factor. In the case of the synchrotron losses, this yields
\begin{equation}
 t_{{\rm syn}, p} \simeq 4.8\times10^{14}\left(\frac{B}{10~{\rm G}}\right)^{-2}\left(\frac{\gamma_p}{100}\right)^{-1}\ [{\rm s}]\,.
\end{equation}
Given the higher energy density of the photon field,  the \ac{ic} cooling time can be up to \(\sim10^4\) times faster. These electrodynamic cooling channels are inefficient as compared to the hardronic mechanisms below. Hereinafter, we do not consider proton \ac{ic}/synchrotron coolings.

The \ac{pp} cooling time can be expressed as
\begin{eqnarray}
\label{eq:t_pp}
 t_{pp} &=& \frac{1}{n_p\sigma_{pp} c \kappa_{pp}},\\ \nonumber
&\simeq& 1.6\times10^6\left(\frac{\taut}{1.1}\right)^{-1}\left(\frac{r_c}{40}\right)\left(\frac{\mbh}{10^8M_\odot}\right)\ [{\rm s}].
\end{eqnarray}
where $\kappa_{pp}\sim 0.5$ is the proton inelasticity of the process and we adopt $\sigma_{pp}=3\times10^{-26}\ {\rm cm}^2$.  Below we adopt the formalism developed by \citet{Kelner2006}. The total cross section of the inelastic \ac{pp} process $\sigma_{pp}$ is represented as a function of the proton energy $E_p=\gamma_p m_pc^2$, 
\begin{eqnarray}
  \sigma_{pp} &\simeq& \\
  &\Big(34.3&+1.88L + 0.25L^2\Big)\left[1-\left(\frac{E_{pp,\rm thr}}{E_p}\right)^4\right]^2~ \rm mb \nonumber
\end{eqnarray}
for $E_p\ge E_{pp,\rm thr}$, where $1\ {\rm mb}=10^{-27}\ {\rm cm}^2$, $L=\log(E_p/1\,\rm TeV)$, and $E_{pp,\rm thr}=1.22$~GeV \citep{Kelner2006}.

The \ac{pg} cooling time via photomeson interactions is
\begin{equation}
\label{eq:t_pg}
 t_{p\gamma}^{-1} = \frac{c}{2\gamma_p^2}
  \int\limits_{\bar{\varepsilon}_{\rm thr}}^{\infty}d\bar{\varepsilon}\sigma_{p\gamma}(\bar{\varepsilon})K_{p\gamma}(\bar{\varepsilon})\bar{\varepsilon} \int\limits_{\bar{\varepsilon}/(2\gamma_p)}^{\infty}d\epsilon\, \frac{U_{\rm ph}(\epsilon)}{\epsilon^4},
\end{equation}
where $\bar{\varepsilon}$ and $\epsilon$ are the photon energy in the proton rest frame and the black hole frame, respectively, 
$U_{\rm ph}$ is the energy density of the photon target, and $\bar{\varepsilon}_{\rm thr} = 145$~MeV. For numerical calculation we follow the formalism suggested by \citet{Kelner2008}.

The \ac{pg} interaction also generates pairs, so-called the Bethe-Heitler pair production process and its cooling timescale is approximated as \citep{Gao2012}
\begin{eqnarray}
t_{\rm BH}^{-1} &\approx&\frac{7(m_{e}c^{2})^{3}\alpha_{f}\sigmat c}{9\sqrt{2}{\pi}m_{p}c^2\gamma_{p}^{2}}\int_{m_ec^2/\gamma_p}^{\infty}d\epsilon\frac{U_{\rm ph}(\epsilon)}{\epsilon^4} \\ \nonumber
&\times& \left\{\left(\frac{2\gamma_{p}\epsilon}{m_ec^2}\right)^{3/2}\left[\log\left(\frac{2\gamma_{p}\epsilon}{m_ec^2}\right) -2/3\right]+2/3\right\},
\end{eqnarray}
where $\alpha_f$ is the fine-structure constant.

\subsection{Acceleration}
In the frame work of \ac{dsa} \cite[e.g.,][]{Drury1983,Blandford1987}, the acceleration time scale can be approximated as 
\begin{equation}
t\DSA\simeq\frac{\eta_{\rm acc}D(E\CR)}{\vsh^2},
\end{equation}
where $D$ is the diffusion coefficient, $E\CR$ is the particle energy, and $\vsh$ is the shock speed. $\eta_{\rm acc}$ is a numerical factor that depends on the shock compression ratio and the spatial dependence of $D$ \citep{Drury1983}. We set $\eta_{\rm acc}=10$. Assuming a Bohm-like diffusion, 
\begin{equation}
D(E\CR )\simeq\frac{\eta_gcE\CR }{3eB},
\end{equation}
where $e$ is the electric charge and $\eta_g$ is the gyrofactor which is the mean free path of a particle in units of the gyroradius. $\eta_g$ characterizes the efficiency of the acceleration. $\eta_g=1$ corresponds to the Bohm limit case. The \ac{dsa} time can be written as
\begin{eqnarray}
&&t\DSA\simeq\frac{10}{3}\frac{\eta_g c R_g}{\vsh^2}, \label{eq:t_acc} \\ \nonumber
&&\simeq 7.6\times10^{-3} \left(\frac{\eta_g}{100}\right)\left(\frac{m_{p/e}}{m_e}\right)\left(\frac{r_c}{40}\right)\left(\frac{B}{10\ {\rm G}}\right)^{-1}\left(\frac{\gamma_{p/e}}{100}\right)\ [{\rm s}].
\end{eqnarray}
where $R_g$ is the gyro radius and $\vsh$ is set as $\vff(R_c)$. $\eta_g$ varies in different astrophysical environments. $\eta_g\sim1$ is possibly seen in a Galactic supernova remnant \citep{Uchiyama2007}, while $\eta_g\sim10^4$ is seen in the case of blazars in the framework of one-zone leptonic models \citep[e.g.,][]{Inoue1996,Finke2008,Inoue2016}.

\section{Properties of Active Supermassive Black Holes}
\label{sec:property}

In this section, we summarize the general observational properties of the central region of \acp{agn} related to high-energy particles in coronae.

\subsection{Broadband Emission from the Core Region}
\label{sec:AGN_SED}
Emission from the \ac{agn} core region mainly arises from two components \citep{Elvis1994}. First is the geometrically thin and optically thick standard accretion disks \citep{Shakura1973}. This standard accretion disk generates a big blue bump from optical to UV attributed by multi-color blackbody radiation. Second is the Comptonized accretion disk photons from the coronal regions above the accretion disk \citep{Katz1976,1977A&A....59..111B,Pozdniakov1977,Sunyaev1980}. This Comptonized emission appears in the X-ray band together with emission reprocessed by the surrounding cold materials, a so-called Compton reflection component \citep[e.g.,][]{Lightman1988,Magdziarz1995,Ricci2011}. 

In this work, for the primary X-ray emission from coronae, we assume a cut-off power-law model in the form of $E^{-\Gamma}\exp(E/E_c)$, where we set $\Gamma=1.9$ and $E_c=300$~keV \citep{Ueda2003,Ueda2014}. For the Compton reflection component, we use the {\tt pexrav} model \cite{Magdziarz1995} assuming a solid angle of $2\pi$, an inclination angle of $\cos i = 0.5$, and the solar abundance for all elements. Since we consider the photons only around the core regions, we ignore the absorption by torus. 

\begin{figure}
 \begin{center}
  \includegraphics[width=9.0cm]{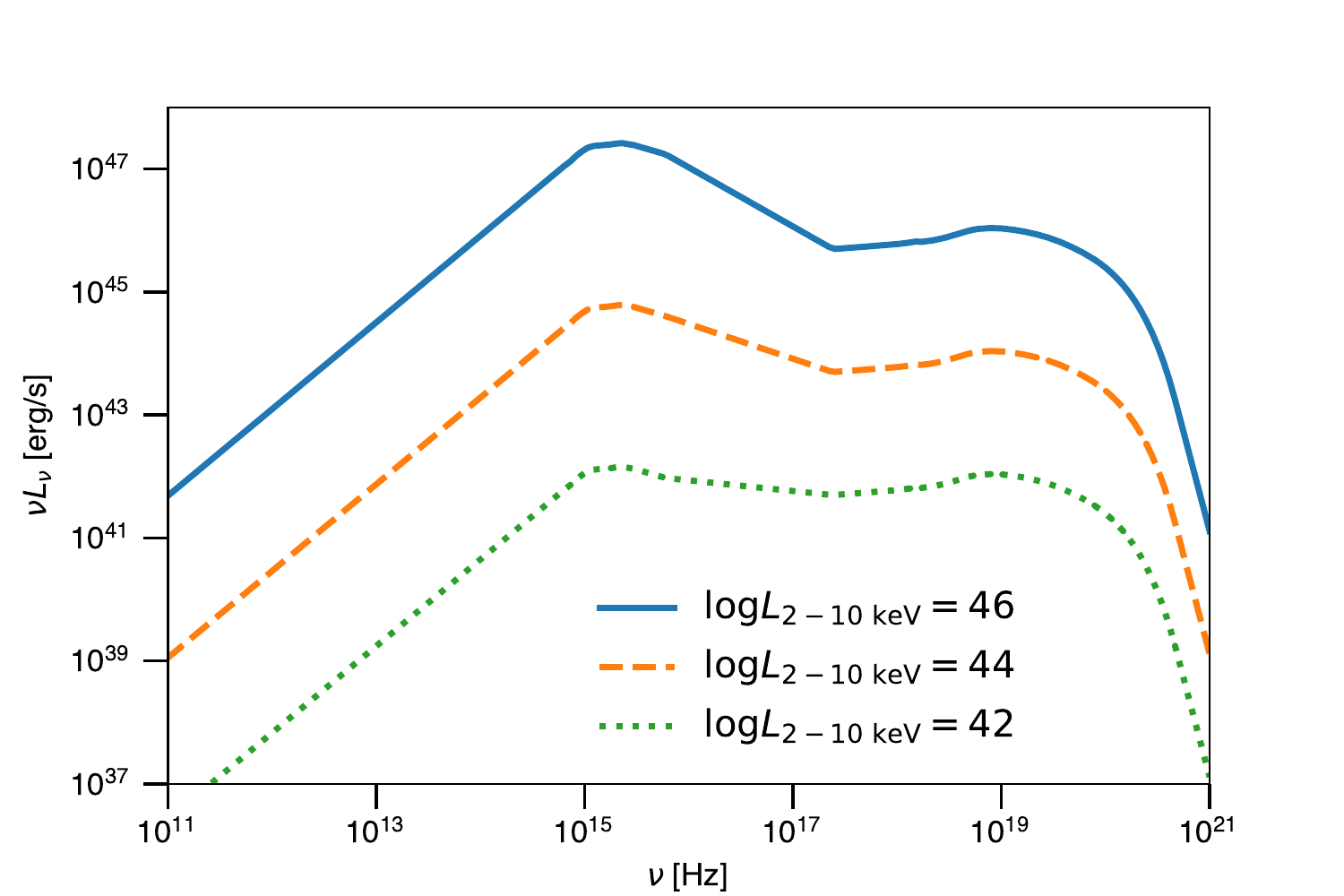}
\caption{The typical broadband spectral energy distribution arising from the core region of \acp{agn}. From top to bottom, each curve corresponds to 2-10~keV luminosity of $10^{46}$, $10^{44}$, $10^{42}~{\rm erg\ s^{-1}}$, respectively. }\label{fig:AGN_SED}
 \end{center}
\end{figure}

The optical-UV accretion-disk \acp{sed} are taken from \citet{Elvis1994}. Here, the primary 2~keV X-ray disk luminosity is connected to the accretion-disk luminosity at 2500~\AA \ as
\begin{equation}
  \log L_{2\ {\rm keV}} = 0.760 \log L_{2500\ \mathrm{\AA}} + 3.508
\end{equation}
based on the study of 545 X-ray selected type~1 \acp{agn} from the XMM-COSMOS survey \citep{Lusso2010}. Between UV and X-ray, following \citet{Lusso2010}, we linearly connect the UV luminosity at 500~\AA \  to the luminosity at 1~keV. Figure~\ref{fig:AGN_SED} shows the broadband \ac{agn} SED arising from the core region for various X-ray luminosities. \ac{agn} core \acp{sed} typically have a spectral peak at $\sim30$~eV corresponding to $\sim10^5$~K (Fig.~\ref{fig:AGN_SED}), which corresponds to the emission radius at around $\sim10R_s$. 

\subsection{Physical Properties of Coronae}
X-ray spectral studies allow us to determine some of the coronal parameters such as the coronal electron temperature $kT_e$ and the Thomson scattering optical depth $\taut$ \citep[e.g.,][]{Brenneman2014}. $k$ is the Boltzmann constant and $T_e$ is the electron temperature in Kelvin. The spectral cutoff at $\sim300$~keV of \ac{agn} core spectra corresponds to the electron temperature of $kT_e\sim100$~keV. The process of Comptonization by thermal plasma is described by the Kompaneets equation \citep{Kompaneets1957}. Here, the photon index of the primary emission is assumed to be 1.9 in this work. This corresponds to $\taut\sim1.1$ based on the solution to the Kompaneets equation \citep{Zdziarski1996} as
\begin{equation}
    \Gamma = \sqrt{\frac{9}{4}+\frac{1}{\theta_e[\taut(\taut+1/3)]}}-\frac{1}{2},
\end{equation}
where the dimensionless electron temperature $\theta_e\equiv kT_e/m_ec^2$. Therefore, in this work, we adopt $kT_e=100$~keV and $\taut=1.1$. These values are consistent with the results from detailed X-ray spectral analysis \cite[e.g.,][]{Fabian2015}.

Recently, utilizing X-ray and radio data, \citet{Inoue2018} found that the coronal magnetic field strength $B$ is approximately $10$~Gauss on scales of $\sim40R_s$ from the \acp{smbh} for two nearby Seyferts whose BH masses are $\sim10^8~M_\odot$\footnote{Contrary to this observational result, recent numerical simulations of the hot accretion flows \citep[e.g.,][]{Kimura2019} shows the magnetic field  enhanced more by the magnetorotational instability \citep[MRI;][]{Balbus1991,Balbus1998}.}. This coronal size is consistent with optical--X-ray spectral fitting studies \citep{Jin2012} and micorolensing observation \citep{Morgan2012}. Thus, in this paper, we set the coronal size as $40R_s$ for all \acp{smbh} and $B=10$~G for $10^8~M_\odot$ \acp{smbh}. 

\citet{Inoue2018} also suggested that the coronae are likely to be advection heated hot accretion flows \citep{Kato2008,Yuan2014} rather than magnetically heated corona \citep{Haardt1991,Liu2002} because the measured magnetic field strength is too weak to keep the coronae hot and is rather consistent with the value based on the self-similar solutions of hot accretion flows \citep{Kato2008,Yuan2014}. Thus, we assume that coronal magnetic field strength scales as
\begin{equation}
B\propto \mbh^{-1/2},	
\end{equation}
following the self-similar solution for the hot accretion flow \citep{Yuan2014} where we ignore dependence on accretion rate and other parameters for simplicity. 

\citet{Mayers2018} have recently investigated a relation between the intrinsic 2--10~keV X-ray luminosity and the mass of central \acp{smbh} using \acp{agn} from the XMM-Newton Cluster Survey. The empirical relation found in \citet{Mayers2018} is given as
\begin{equation}
\mbh=2\times10^7 M_\odot \left[\frac{L_{2-10\ {\rm keV}}}{1.155\times10^{43}~{\rm erg\ s^{-1}}}\right]^{0.746}.
\end{equation}
Using this relation, we can convert X-ray luminosities to masses of central \acp{smbh}.

\subsection{Internal Gamma-ray Attenuation in Coronae}
\label{subsec:gg_int}
Accelerated electrons and protons in coronae would emit gamma rays (see \S \ref{sec:AGN_SED}). However, high energy gamma-ray photons are attenuated by photon-photon pair production interactions ($\gamma\gamma \rightarrow e^+e^-$) with low-energy photons. For isotropic target photons the pair production cross section achieves its maximum of \(\approx0.2\sigmat\) when a gamma-rays of energy $E_\gamma$ interacts with a low-energy photon with energy \citep[see, e.g.,][]{Aharonian_book}
\begin{equation}
\label{eq:ene_ebl}
\epsilon_{\rm peak}\simeq \frac{3.5m_e^2c^4}{E_\gamma}\simeq1\left(\frac{1{\rm \ TeV}}{E_\gamma}\right)\ {\rm eV}.
\end{equation}
In terms of wavelength, $\lambda_{\rm peak}\simeq1.4(E_\gamma[{\rm TeV}])\ \mu{\rm m}$.

\begin{figure}
 \begin{center}
  \includegraphics[width=9.0cm]{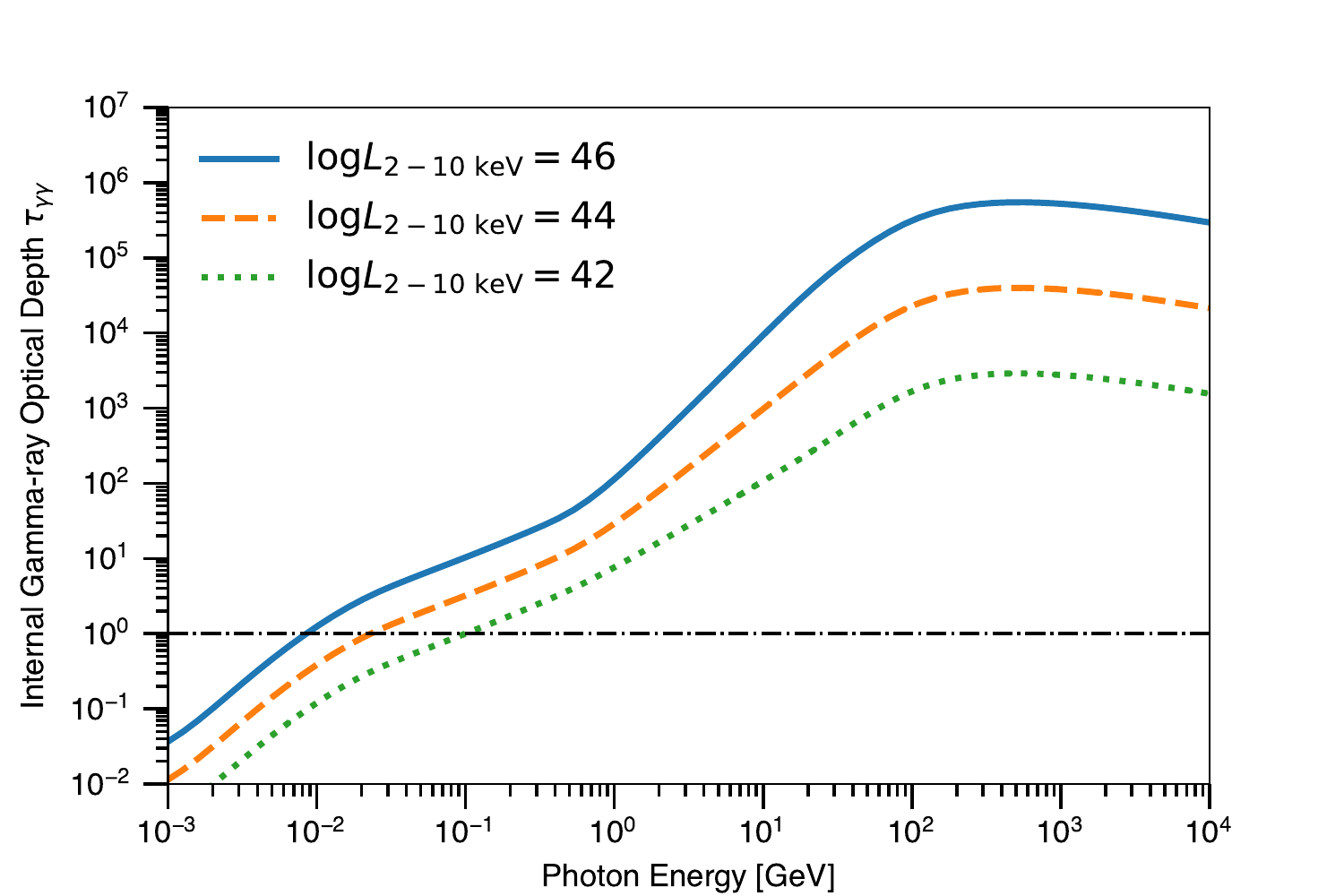}
\caption{Internal gamma-ray optical depth in the core region of \acp{agn}. From top to bottom, each curve corresponds to 2-10~keV luminosity of $10^{46}$, $10^{44}$, $10^{42}~{\rm erg\ s^{-1}}$, respectively. The horizontal dot-dashed line represents $\tau_{\gamma\gamma}=1$.}\label{fig:tau_gg}
 \end{center}
\end{figure}

\begin{figure*}[tb!]
 \begin{center}
  \includegraphics[width=17cm]{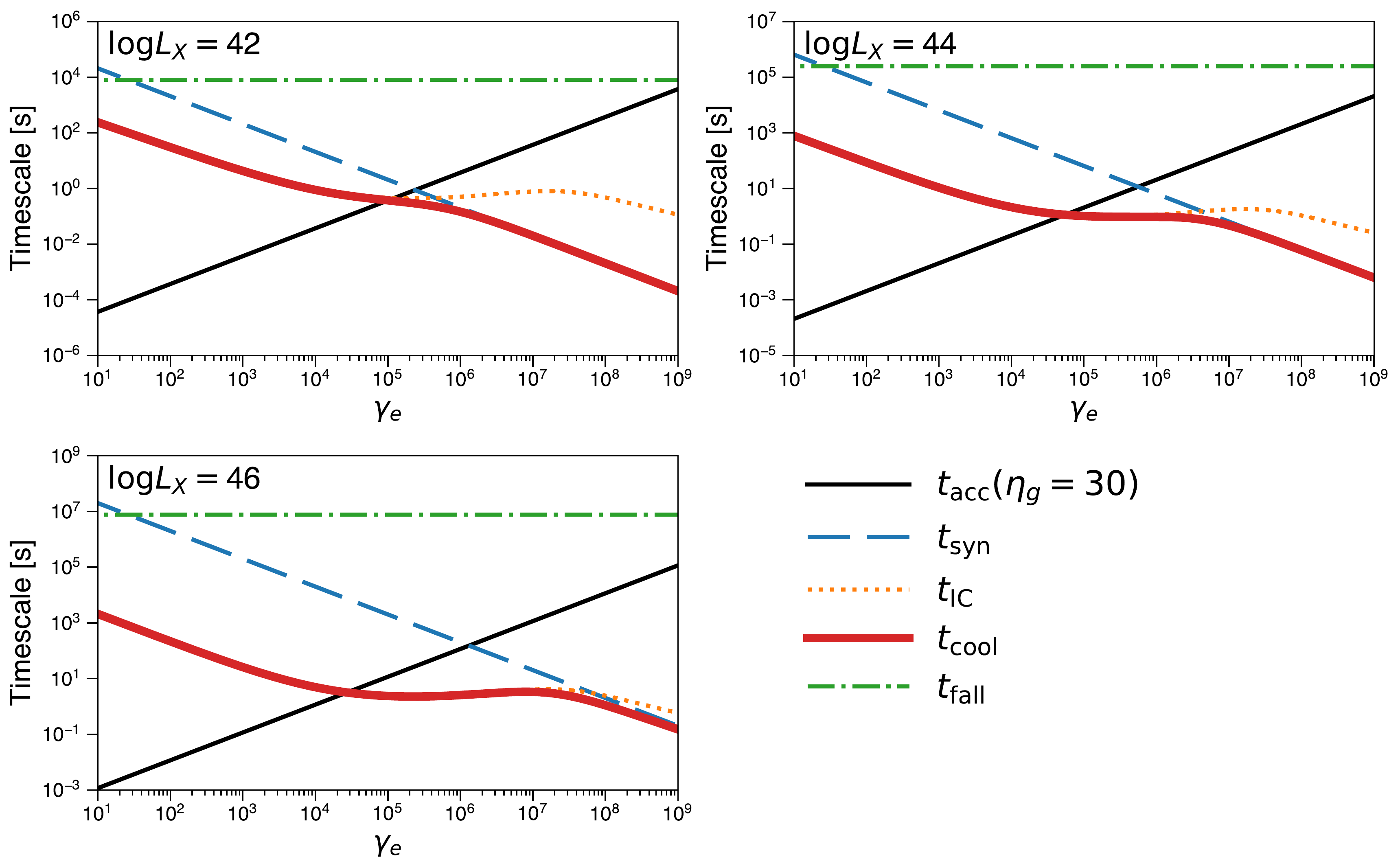}
\caption{Electron energy losses in \ac{agn} coronae together with acceleration and dynamical timescales. Each panel corresponds to different 2--10~keV X-ray luminosity as indicated in panels. Thin solid line shows the acceleration timescale assuming DSA. Dashed, dotted, and thick solid curve corresponds to synchrotron cooling, \ac{ic} cooling, and total cooling timescale, respectively. Dot-dashed curve shows the free-fall timescale. In these plots, we set $\taut=1.1$, $R_c=40R_s$, $kT_e=100$~keV, and $\eta_g=30$. We note that the vertical axis ranges are different in each panel.}\label{fig:time_electron}
 \end{center}
\end{figure*}

Abundant photons are emitted from the \ac{agn} core region (Fig.~\ref{fig:AGN_SED}). From the SED of \ac{agn} core regions as given in \S~\ref{sec:AGN_SED},
we can compute the optical depth for high-energy gamma rays to $\gamma\gamma$ pair production interactions. The cross section for this process is \citep{1934PhRv...46.1087B,Heitler1954}
\begin{eqnarray}
&&\sigma_{\gamma\gamma}(E_\gamma,\epsilon,\theta)=\frac{3\sigmat }{16}(1-\beta^2) \nonumber \\
&&\times\left[2\beta(\beta^2-2)+(3-\beta^4)\ln\left(\frac{1+\beta}{1-\beta}\right)\right],
\end{eqnarray}
where $\beta$ is 
\begin{equation}
\beta\equiv\sqrt{1-\frac{2m_e^2c^4}{\epsilon E_\gamma(1-\mu)}};\ \ \mu\equiv\cos\theta.
\end{equation}
where $\theta$ is the angle between the colliding photons' momenta. 

For a photon with an energy of $E_\gamma$, the $\gamma\gamma$ optical depth is
\begin{equation}
\label{eq:tau_gg}
\tau_{\gamma\gamma}(E_\gamma)=\int\limits_{-1}^{1}d\mu\int\limits_{\epsilon_{\rm th}}^{\infty}d\epsilon\frac{1-\mu}{2} \frac{U_{\rm ph}(\epsilon)}{\epsilon^2}\sigma_{\gamma\gamma}(E_\gamma,\epsilon,\theta)R_c 
\end{equation}
where $\epsilon_{\rm th}$ is the pair production threshold energy,
\begin{equation}
\epsilon_{\rm th}=\frac{2m_e^2c^4}{E_\gamma(1-\mu)}.
\end{equation}
Integration over the interaction angle in Eq.~\eqref{eq:tau_gg} can be performed analytically resulting in the angle averaged \(\gamma\gamma\) cross section \citep{Aharonian_book}:
\begin{eqnarray}
\sigma_{\gamma \gamma}  & =&
\frac{3 \sigmat}{2 s^2}
\left[ \left(s+ \frac{1}{2} \ln s- \frac{1}{6} +\frac{1}{2s} \right)
\ln(\sqrt{s}+ \sqrt{s-1})  -  \right. 
\nonumber \\
 & & \left.  
\left(s+ \frac{4}{9} - \frac{1}{9s}\right)  \sqrt{1- \frac{1}{s}}\right] \,,
\end{eqnarray}
where \(s=E_\gamma \epsilon/m_e^2 c^4\). 

Figure~\ref{fig:tau_gg} shows the internal gamma-ray optical depth in the core region for various X-ray luminosities. The core region is expected to be optically thick against gamma-ray photons above 10--100~MeV depending on disk luminosities. Such high optical thicknesses against pair production in \ac{agn} coronae are well known
\citep[e.g.,][]{1971MNRAS.152...21B,Done1989,Fabian2015} based on the compactness parameter argument \citep{Guilbert1983}.

\section{Timescales} 
\label{sec:timescales}
Given the observed properties of AGN core regions, we can estimate the various timescales of high energy particles in the coronae. Figure~\ref{fig:time_electron} shows the cooling rates of electrons in the coronae for different energy-loss processes, together with the acceleration rate and the free-fall timescale following \S~\ref{sec:acceleration} and parameters presented in \S~\ref{sec:property}. We set $\eta_g=30$ in the figure, which reproduces the IceCube neutrino background fluxes as discussed later in \S~\ref{sec:background}. Each panel corresponds to 2-10~keV X-ray luminosity of $10^{42}$, $10^{44}$, $10^{46}~{\rm erg\ s^{-1}}$. 

Due to the intense broadband radiation field, the cooling is dominated by the Compton cooling. However, at higher energy regions, the main cooling channel is replaced by synchrotron cooling because of the \ac{kn} effect. The more luminous AGNs tend to have more efficient \ac{ic} cooling effect, as the target photon density increases. When we assume $\eta_g=30$, electron acceleration up to $\gamma_e\sim10^5$ ($\sim50$~GeV) is feasible in \ac{agn} coronae at various luminosities. Therefore, synchrotron radiation through coronal magnetic fields and gamma-ray emission by Comptonization of disk photons are naturally expected in AGN coronae.

\ac{alma} spectra of two nearby Seyferts, whose X-ray luminosities are about $10^{44}\ {\rm erg\ s^{-1}},$ extends their radio synchrotron power-law spectra at least up to 230~GHz, which corresponds to $\gamma_e\sim80$ given the magnetic field strength of $10$~G \citep{Inoue2018}\footnote{This frequency limit is due to the instrumental coverage of the \ac{alma} band-6 receiver. Therefore, the emission itself is likely to extend to higher frequencies, even though those emission signals would be buried in thermal dust emission.}. As shown in the top right panel (the case of $\log L_X=44$) in Fig. \ref{fig:time_electron}, relativistic electrons with $\gamma_e\sim80$ seen by \ac{alma} can be easily accelerated in AGN coronae. Notably, such electrons can be accelerated even by a low efficiency acceleration process, e.g., with $\eta_g\sim10^6$. For this energy, Compton cooling is the dominant energy loss process. As the cooling timescale for $\gamma_e\sim80$ is about 100~s, flux variability in the radio synchrotron emission is expected, some Seyferts are already known to show a flux variation at least in day scales \citep{Baldi2015}. Further dense light curve observations may see shorter timescale variabilities.

\begin{figure*}
 \begin{center}
  \includegraphics[width=17cm]{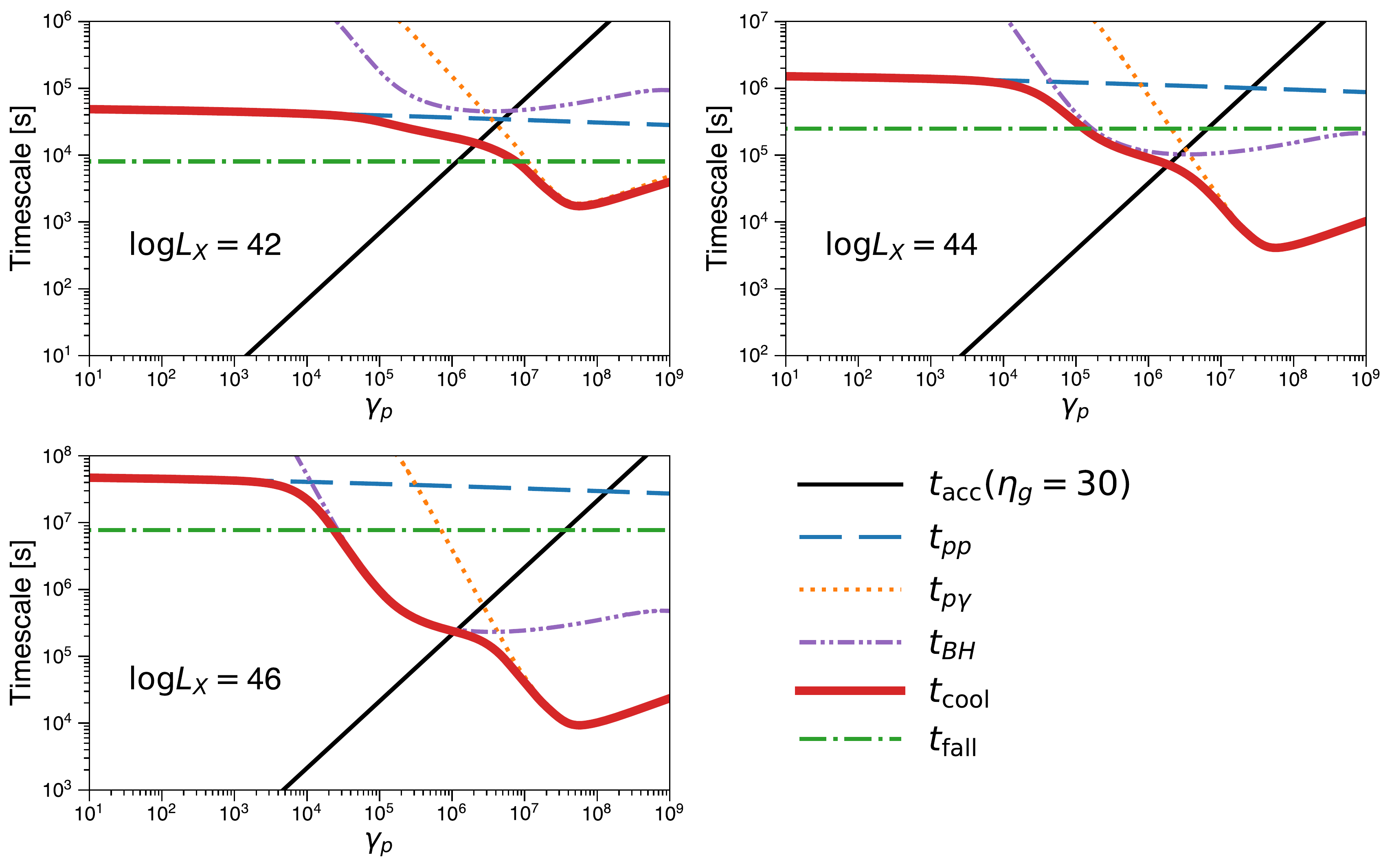}
\caption{Same as in Fig.~\ref{fig:time_electron}, but for protons. Dashed, dotted, double-dot-dashed, and thick solid curve corresponds to \ac{pp} cooling,  \ac{pg} cooling, BH cooling, and total cooling timescale, respectively.}\label{fig:time_proton}
 \end{center}
\end{figure*}

Similar to Fig.~\ref{fig:time_electron} for electrons, Fig.~\ref{fig:time_proton} shows the timescales for high energy for various luminosities. As in Fig.~\ref{fig:time_electron}, we set $\eta_g=30$. Since synchrotron and Compton cooling are not effective for protons in our case, we do not show these timescales in the figure. 

It is evident that protons can be accelerated up to $\gamma_p\sim10^6$ ($\sim1$~PeV) in \ac{agn} coronae for various luminosities. Maximum attainable energy is controlled by different processes for different luminosity AGNs due to SED and size dependence. For low-luminosity Seyferts ($L_X<10^{44}\ {\rm erg\ s^{-1}}$), acceleration is limited by the dynamical timescale rather than radiative cooling, while it becomes limited by the Bethe-Heitler cooling for higher luminosity objects. As the luminosity increases, \ac{pg} and Bethe-Heitler cooling effects become more prominent. At higher luminosities, the Bethe-Heitler processes dominate the energy loss process of high energy particles. Therefore, in cases of high luminosity objects, resulting hadronic gamma-ray and neutrino spectra in the TeV band will show spectral suppression due to the Bethe-Heitler processes \citep[see e.g.,][for the cases of gamma-ray burst]{Murase2008}.

\section{Particle Spectrum}
\label{sec:particle_spectrum}
The steady state particle distributions $n=dN/d\gamma$ can be derived from the solution of the transport equation \citep{Ginzburg1964}
\begin{equation}
    \frac{\partial}{\partial \gamma}\left(\dot{\gamma}_{\rm cool}n\right)+\frac{n}{t_{\rm fall}} = Q(\gamma),
\end{equation}
where $\dot{\gamma}_{\rm cool}$ is the total cooling rate, $Q(\gamma)$ is the injection function, which describes phenomenologically some acceleration process, e.g., \ac{dsa}. The injection function for non-thermal protons and electrons is set as $Q(\gamma) = Q_0\gamma^{-p_{\rm inj}}\exp(-\gamma/\gamma_{\rm max})$. Here, $\gamma_{\rm max}$ is the maximum Lorentz factor determined by balancing the acceleration and cooling time scales (Figures. \ref{fig:time_electron} and \ref{fig:time_proton}). The corresponding solution is
\begin{equation}
\label{eq:electron_spectrum}
    n=\frac{1}{\dot{\gamma}_{\rm cool}}\int\limits_\gamma^{\infty}Q(\gamma')e^{-T(\gamma,\gamma')} d\gamma',
\end{equation}
where
\begin{equation}
    T(\gamma_1, \gamma_2) = \frac{1}{t_{\rm fall}}\int\limits_{\gamma_1}^{\gamma_2}\frac{d\gamma}{\dot{\gamma}_{\rm cool}}
\end{equation}
By solving Equation. \ref{eq:electron_spectrum}, we obtain a steady-state spectrum of the non-thermal particles. 

\begin{figure}
 \begin{center}
  \includegraphics[width=9.0cm]{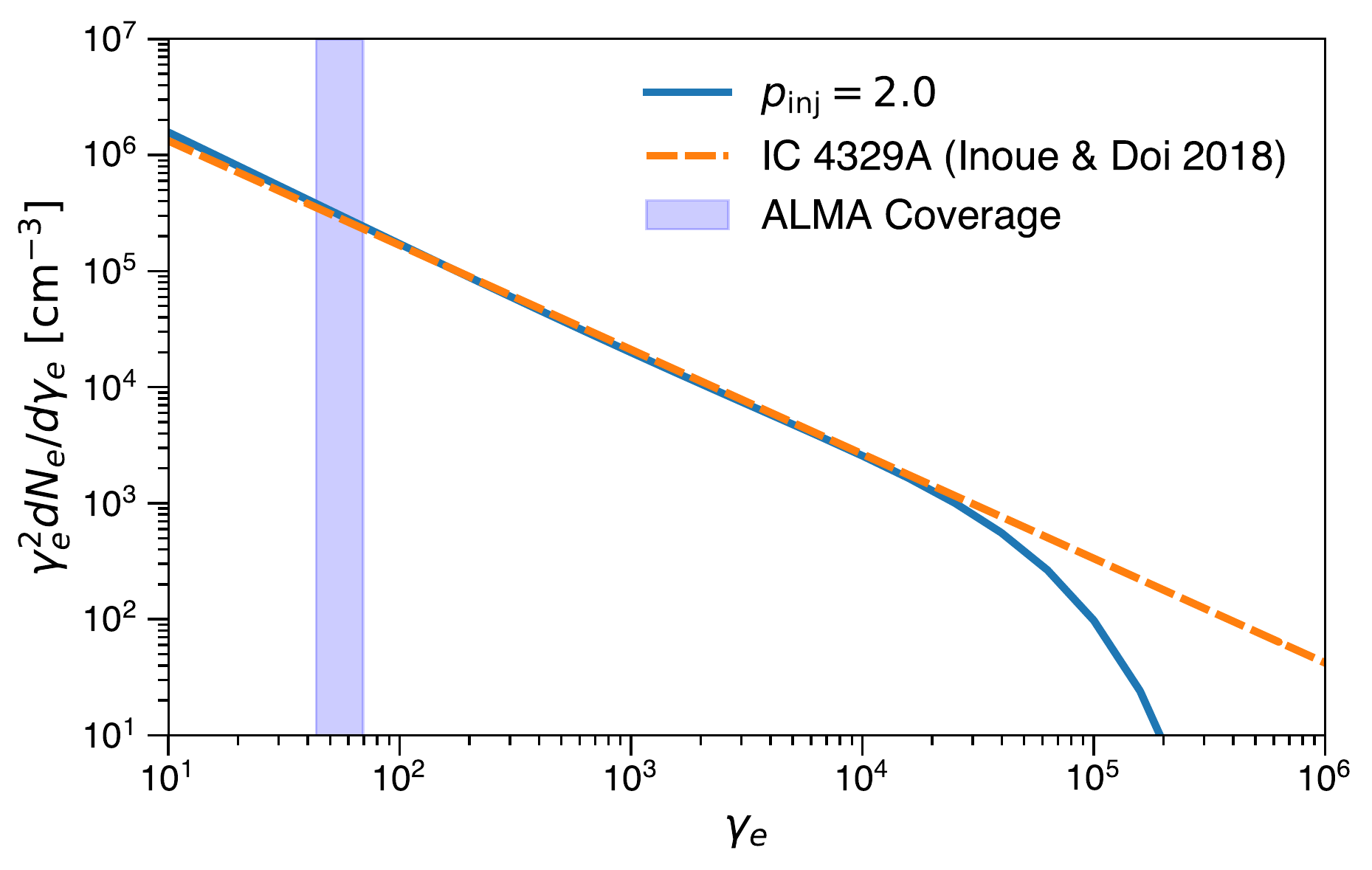}
\caption{The steady-state electron spectral distribution in \ac{agn} coronae. Solid curve corresponds to the model with $p_{\rm inj}=2.0$. We set $\mbh=10^8M_\odot$, $r_c = 40$, $B=10$~G, $kT_e=100$~keV, $\taut=1.1$, and $\eta_g=30$. Dashed curve corresponds to the observationally determined electron distribution for IC~4329A \citep{Inoue2018}. The shaded region shows the Lorentz factors responsible for the observed radio spectrum.}\label{fig:electron_spectrum}
 \end{center}
\end{figure}

Fig.~\ref{fig:electron_spectrum} shows the steady-state non-thermal electron spectrum obtained for the injection spectral index of $p_{\rm inj}=2.0$ together with the observationally determined electron spectral distribution for IC~4329A \citep{Inoue2018}. \ac{alma} observed non-thermal synchrotron radiation between 90.5~GHz and 231~GHz which corresponds to the electron Lorentz factors between 50 and 80, respectively.  The corresponding region is shown as the shaded region in the Fig.~\ref{fig:electron_spectrum}.

For the calculation of the steady-state spectrum, we set $\mbh=10^8M_\odot$, $r_c = 40$, $B=10$~G, $kT_e=100$~keV, $\taut=1.1$, and $\eta_g=30$. The synthetic electron distribution obtained for $p_{\rm inj}=2.0$ nicely reproduces the observationally determined electron spectrum in the energy range constrained by the observations. This injection index is  naturally expected in a simple \ac{dsa} scenario for a strong shock. 

The resulting particle spectrum at $\gamma_e>10^4$ becomes softer than observationally determined index at $50\lesssim\gamma_e\lesssim80$. This is because of the influence of the cutoff imposed by the particle cooling. Therefore, if we consider the high energy synchrotron or \ac{ic} spectral shapes, the cooling effects should be taken into account accurately. Even though the electron spectrum extends down to lower energies, it is hard to see the corresponding synchrotron emission due to synchrotron self-absorption effect \citep{Inoue2014}.

The calculated electron spectrum is renormalized to agree with the observationally determined spectrum, which is achieved if the non-thermal electrons contains $f_{\rm nth}=0.03$ of the energy in thermal leptons. We note that, in order to define the energy content in the non-thermal particles, we formally integrate above $\gamma_e=1$ in this study. We keep this fraction for non-thermal electron energy fixed in calculations below for all Seyferts. 

The energy fraction of non-thermal electrons was fixed to $\xi_{\rm nth}=0.04$ in \citet{Inoue2018}. $\xi_{\rm nth}$ is defined beyond the break electron Lorentz factor, while $f_{\rm nth}$ is above $\gamma_e=1$. That amount of non-thermal electrons overproduces the MeV background flux given the measured electron spectral index (see \S. \ref{sec:background}). To be consistent with the observed cosmic MeV gamma-ray background flux, we set $\xi_{\rm nth}=0.015$ in this work, which corresponds to $f_{\rm nth}=0.03$. The obtained best fit parameters with this fraction for the radio spectrum of IC~4329A is $p=2.9\pm0.9$, $B=11.4\pm5.6$~G, and $r_c = 42.7\pm7.8$, which are very similar to those obtained for the case of $\xi_{\rm nth}=0.04$. We adopt these parameters for the observationally determined electron distribution in the Fig. \ref{fig:electron_spectrum}. Fitting results for the other parameters were also the same as those with $\xi_{\rm nth}=0.04$.

Here, the total shock power $\Psh$ can be estimated as 
\begin{eqnarray}
	\Psh&=&4\pi R_c^2 n_pm_p v^3_{\rm sh}/2\\ \nonumber
	&\simeq&2.2\times10^{45}\left(\frac{\taut}{1.1}\right)\left(\frac{r_c}{40}\right)^{-1/2}\left(\frac{\mbh}{10^8M_\odot}\right)\ {\rm erg\ s^{-1}}.
\end{eqnarray}
For objects with $L_{\rm X}=10^{44}~{\rm erg\ s^{-1}}$, $f_{\rm nth}=0.03$ corresponds to $\sim5$\% of the shock power is injected into acceleration of electrons. This high value implies that if \ac{dsa} is responsible for particle acceleration in \ac{agn} coronae then processes regulating injection of electrons into \ac{dsa} are very efficient. For example in the case of \ac{dsa} in supernovae remnants non-thermal electrons obtain only $\sim1$\% of energy transferred to non-thermal protons \citep{Ackermann2013}. Detailed consideration of the reasons of this unusually high efficiency of electron acceleration is beyond the scope of this paper, however we note that a significant presence of positrons may affect the ratio \citep[see, e.g.,][]{2015PhRvL.114h5003P}.  Given these uncertainties, for protons we set that the same energy injection rate is achieved as for electrons. This power appears to be sufficient to explain the observed IceCube neutrino fluxes.

For the other object, NGC~985, the observed electron spectral index is $2.11\pm0.28$ \citep{Inoue2018}, which is hard considering the radiative cooling effect. 
Cascade components would have such a hard spectrum below the threshold energy \citep[see, e.g.,][]{2003APh....19..525A}. In addition, due to the quality of data at low frequencies, we could not precisely determine the other components such as free-free emission and synchrotron emission from star formation activity, and synchrotron emission from the jet. Those uncertainties may resulted in a less reliable measurement of the corona emission spectrum slope. Further observations are required to determine the radio spectral properties in NGC~985 precisely.

\begin{figure*}
 \begin{center}
  \includegraphics[width=8.9cm]{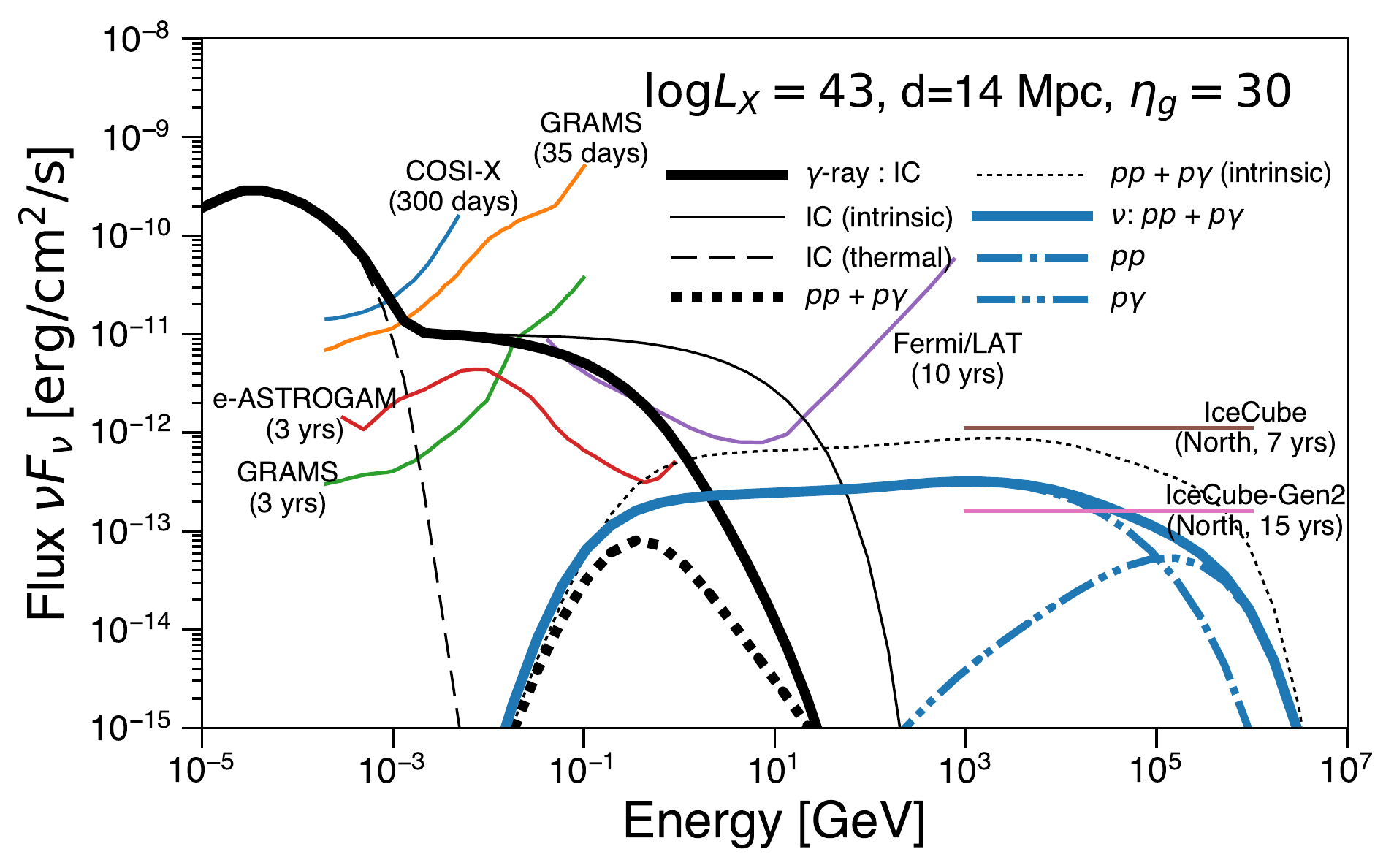}
  \includegraphics[width=8.9cm]{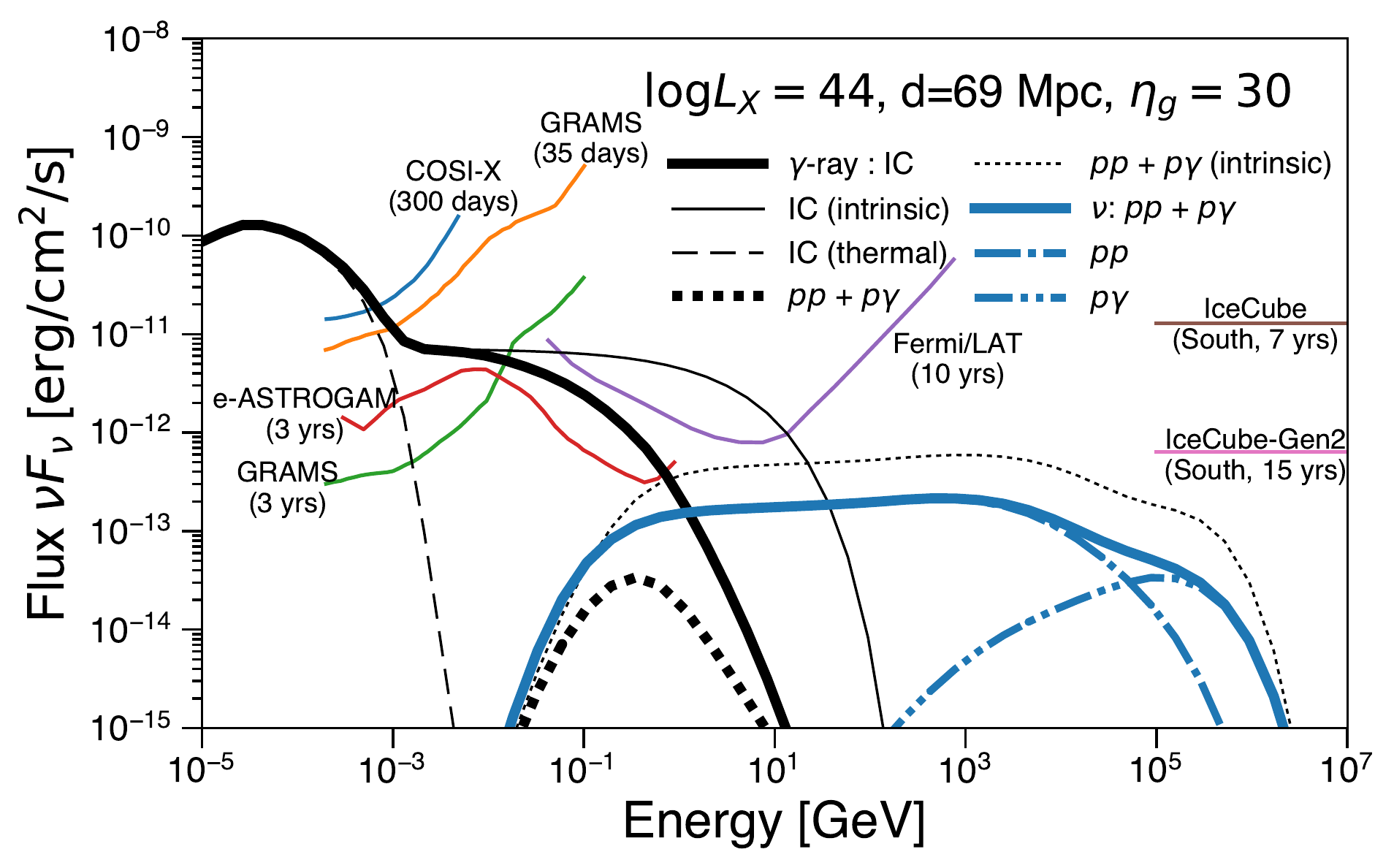}
\caption{{\it Left}: Gamma-ray and neutrino spectrum per flavour from an \ac{agn} coronae with $p_{\rm inj}=2.0$ and $\eta_g=30$. We set 2-10~keV luminosity of $10^{43}~{\rm erg\ s^{-1}}$ at a distance of 14~Mpc, which roughly corresponds to NGC~4151. We renormalize the overall fluxes in order to match the {\it Swift}/BAT flux of NGC~4151 at 14-195~keV \citep{Oh2018}. The thick black solid and thick dot curve shows gamma rays from \ac{ic} interaction and \ac{pp}+\ac{pg} interaction including internal and EBL attenuation effect. Each thin curve shows the spectrum before the attenuation. The black dashed curve shows the \ac{ic} spectrum considering only thermal electrons, in which the effect of reflection is taken into account. The blue dot-dashed, double-dot-dashed, and solid curve shows the neutrino contribution per flavour of \ac{pp} interaction, \ac{pg} interaction, and the sum of the two, respectively. The non-thermal electrons in coronae are assumed to carry 3\% of the total lepton energies. We assume the injection powers in electrons and protons are the same. For the comparison, we overplot the sensitivity curve of {\it COSI-X} (300~days), {\it e-ASTROGAM} \citep[3~yrs;][]{DeAngelis2017}, {\it GRAMS} \citep[35~days;][]{Aramaki2019}, {\it GRAMS} \citep[3~yrs;][]{Aramaki2019}, and {\it Fermi}/LAT (10~yrs). We also plot the sensitivity of IceCube and IceCube-Gen2 at $\delta=30^\circ$ \citep{vanSanten2017}. {\it Right:} The same as the {\it Left} panel, but we set 2-10~keV luminosity of $10^{44}~{\rm erg\ s^{-1}}$ at a distance of 69~Mpc which roughly corresponds to IC~4329A. We renormalize the overall fluxes in order to match the {\it Swift}/BAT flux of IC~4329A at 14-195~keV \citep{Oh2018}. For the IceCube sensitivity, we show that at $\delta=-30^\circ$.}\label{fig:SED_gamma_nu}
 \end{center}
\end{figure*}

\section{Gamma Rays and Neutrinos from AGN Coronae}
\label{sec:g_nu_AGN}

Accelerated electrons and protons in AGN coronae generate gamma-ray and neutrino emission through \ac{ic} scattering, $pp$ interaction, and $p\gamma$ interaction. Adopting a steady-state particle spectrum, we calculate the resulting gamma-ray and neutrino spectra from AGN coronae. We follow \citet{Blumenthal1970} for the gamma-ray emission due to the \ac{ic} scattering by non-thermal electrons. We calculate the gamma-ray and neutrino emission induced by hadronic interactions following \citet{Kelner2006} for \ac{pp} interactions and \citet{Kelner2008} for \ac{pg} interactions. For simplicity, we do not take into account \ac{ic} scattered emission by secondary electrons and positrons. For the thermal Comptonization spectra, we adopt the AGN SED shown in Fig.~\ref{fig:AGN_SED} which takes into account reflection components but does not account for attenuation by torus. The torus attenuation is mainly relevant for $\lesssim30$~keV, which is below the range of our interest.

Figure~\ref{fig:SED_gamma_nu} shows the resulting gamma-ray and neutrino spectra for two cases. The neutrino flux is shown in the form of per flavour. The left panel of the figure shows the case with a 2-10~keV luminosity of $10^{43}~{\rm erg\ s^{-1}}$ at a distance of 14~Mpc, while the right panel shows the case with a luminosity of $10^{44}~{\rm erg\ s^{-1}}$ at a distance of 69~Mpc. The former and the latter roughly corresponds to NGC~4151 and IC~4329A, respectively. NGC~4151 is the brightest Seyfert in the X-ray sky \citep{Oh2018}. For the comparison, the overall fluxes of both panels are renormalized to match with the {\it Swift}/BAT flux of NGC~4151 and IC~4329A, respectively, at 14-195~keV \citep{Oh2018}.  We note that we do not calculate the detailed X-ray spectra of each objects, which is beyond the scope of this paper. 

We set the injection spectral index of $p_{\rm inj}=2.0$ and the gyrofactor of $\eta_g=30$ for both electrons and protons (See \S~\ref{sec:particle_spectrum}). We also set the same injection power into protons and electrons as described in \S. \ref{sec:particle_spectrum}. The target photon density for \ac{ic} scatterings and \ac{pg} interactions is defined as $U_{\rm ph}(\epsilon)$ (See \S~\ref{sec:AGN_SED}). Since we assume a uniform spherical source, gamma-ray photons are attenuated by internal photon field by a factor of $3u_{\rm int}(\tau_{\rm int})/\tau_{\rm int}$, where $u_{\rm int}(\tau)=1/2 + \exp(-\tau)/\tau - [1 - \exp(-\tau)]/\tau^2$ \citep[See Sec. 7.8 in][]{Dermer2009}, where $\tau_{\rm int}$ is the internal gamma-ray optical depth (See \S. \ref{subsec:gg_int}). Gamma rays are also attenuated by the \ac{ebl} during the propagation in the intergalactic space. We adopt \citet{Inoue2013} for the \ac{ebl} attenuation. 

For the comparison, we also show the expected sensitivity curve of planned MeV missions: {\it COSI-X} (300~days)\footnote{{\it COSI} collaboration website (The Compton Spectrometer and Imager \url{http://cosi.ssl.berkeley.edu/}}, {\it e-ASTROGM} \citep[3~yrs,][]{DeAngelis2017}\footnote{{\it e-ASTROGAM} collaboration website (enhanced ASTROGAM \url{http://eastrogam.iaps.inaf.it/}}, {\it GRAMS} \citep[35~days,][]{Aramaki2019}, and {\it GRAMS} \citep[3~yrs,][]{Aramaki2019}. 10-yr sensitivity of {\it Fermi}/LAT\footnote{{\it Fermi}/LAT collaboration website (The Large Area Telescope \url{http://www.slac.stanford.edu/exp/glast/groups/canda/lat_Performance.htm}} is also shown. We also plot the sensitivity of neutrino detectors: IceCube\footnote{IceCube collaboration website (\url{https://icecube.wisc.edu/}} and IceCube-Gen2 \citep{vanSanten2017}. For the left panel, we assume the declination $\delta$ of $30^\circ$, while $-30^\circ$ for the right panel.

Since the spectral index of electrons is $\sim3$ after radiative cooling, the resulting non-thermal gamma-ray spectrum is flat in $\nu F_\nu$ in the MeV band which appears after the thermal cutoff. Given the cooling limited maximum energy $\gamma_e\sim10^5$, the intrinsic \ac{ic} spectrum can extend up to $\sim100$~GeV. However, due to the strong internal gamma-ray attenuation effect, the spectra will have a cutoff around 100~MeV in both cases. In the sub-MeV band, the spectrs show super-thermal tails due to the combination of thermal and non-thermal components and a spectral hardening at $\sim1$~MeV. These superthermal and flat spectral tails should be tested by future MeV gamma-ray missions. Ballon flights with such as {\it GRAMS} \citep{Aramaki2019} and {\it SMILE} \citep{Takada2011,Komura2017}\footnote{{\it SMILE} collaboration website (The Sub-MeV gamma-ray Imaging Loaded-on-balloon Experiment \url{http://www-cr.scphys.kyoto-u.ac.jp/research/MeV-gamma/wiki/wiki.cgi?page=Top_en}} may be able to catch this superthermal tail. And, satellite-class MeV missions such as {\it e-ASTROGAM} \citep{DeAngelis2017}, {\it AMEGO}\footnote{{\it AMEGO} collaboration website (The All-sky Medium Energy Gamma-ray Observatory \url{https://asd.gsfc.nasa.gov/amego/}}, and {\it GRAMS} \citep{Aramaki2019} will be able to see also the non-thermal power-law tail. For the case of NGC~4151, {\it Fermi}/LAT may be able to see the signature with its 10~yrs survey. However, the expected flux is almost at the sensitivity limit. Thus, it may need further exposures for {\it Fermi}/LAT to see the coronal emission.

The \ac{pp} and \ac{pg} production efficiency is given by the ratio between the dynamical timescale (Eq. \ref{eq:t_fall}) and the interaction timescales (Eqs. \ref{eq:t_pp} and \ref{eq:t_pg}). The \ac{pp} production efficiency is analytically given as
\begin{equation}
f_{pp}=\frac{t_{\rm fall}}{t_{pp}} \simeq 0.16 \left(\frac{\taut}{1.1}\right)\left(\frac{r_c}{40}\right)^{-1/2}.
\end{equation}
Gamma rays and neutrinos induced by hadronic interactions carry $1/3$ and $1/6$ of those interacted hadron powers. Therefore, hadronic gamma-ray and neutrino luminosity is expected to be $\sim5$\% and $\sim3$\% of the intrinsic proton luminosity, respectively. Since we assume the same energy injection to electrons and protons and the coronal Thomson scattering optical depth is 1.1, before the attenuation, we have hadronic gamma-ray and neutrino fluxes are $\sim5$\% and $\sim3$\% of the \ac{ic} gamma-ray fluxes.

The \ac{pp} and \ac{pg} induced gamma rays are also mostly attenuated by the internal photon fields. Thus, we do not expect any $\gtrsim$GeV gamma-ray emission from Seyferts. Moreover, the intrinsic gamma-ray energy fluxes due to hadronic interactions is about a factor of 10 less than that by primary electrons because of radiative efficiency differences between protons and electrons. This implies that gamma rays produced by secondary pairs should not significantly alter the resulting spectra. Therefore, we can safely ignore the cascade contribution. 

On the contrary to gamma rays, neutrinos induced by hadronic interactions can escape from the system without any attenuation. Since we adopt the same $p_{\rm inj}=2$ for protons as for electrons, we expect a flat $\nu F_\nu$ spectrum for neutrinos, to which \ac{pp} makes dominant contribution. At higher energies, especially in the case of IC~4329A, \ac{pp} and \ac{pg} spectra are suppressed due to the Bethe-Heitler cooling process. The exact position of the cutoff energy depends on the assumed $\eta_g$. Here, as described later, we set $\eta_g=30$ in order to be consistent with the IceCube background flux measurements. This gyrofactor results in a neutrino spectral cutoff around 100~TeV. Although it is difficult to see neutrino signals from individual Seyferts with the current generation of IceCube, it would be possible to see bright Seyferts in the northern hemisphere in the era of IceCube-Gen2 \citep[see also][for more general arguments]{Murase2016}. Therefore, even though Seyferts are faint in the GeV gamma-ray band, future MeV gamma-ray and TeV neutrino observations can test our scenario.

\begin{figure*}
 \begin{center}
	\includegraphics[width=17cm]{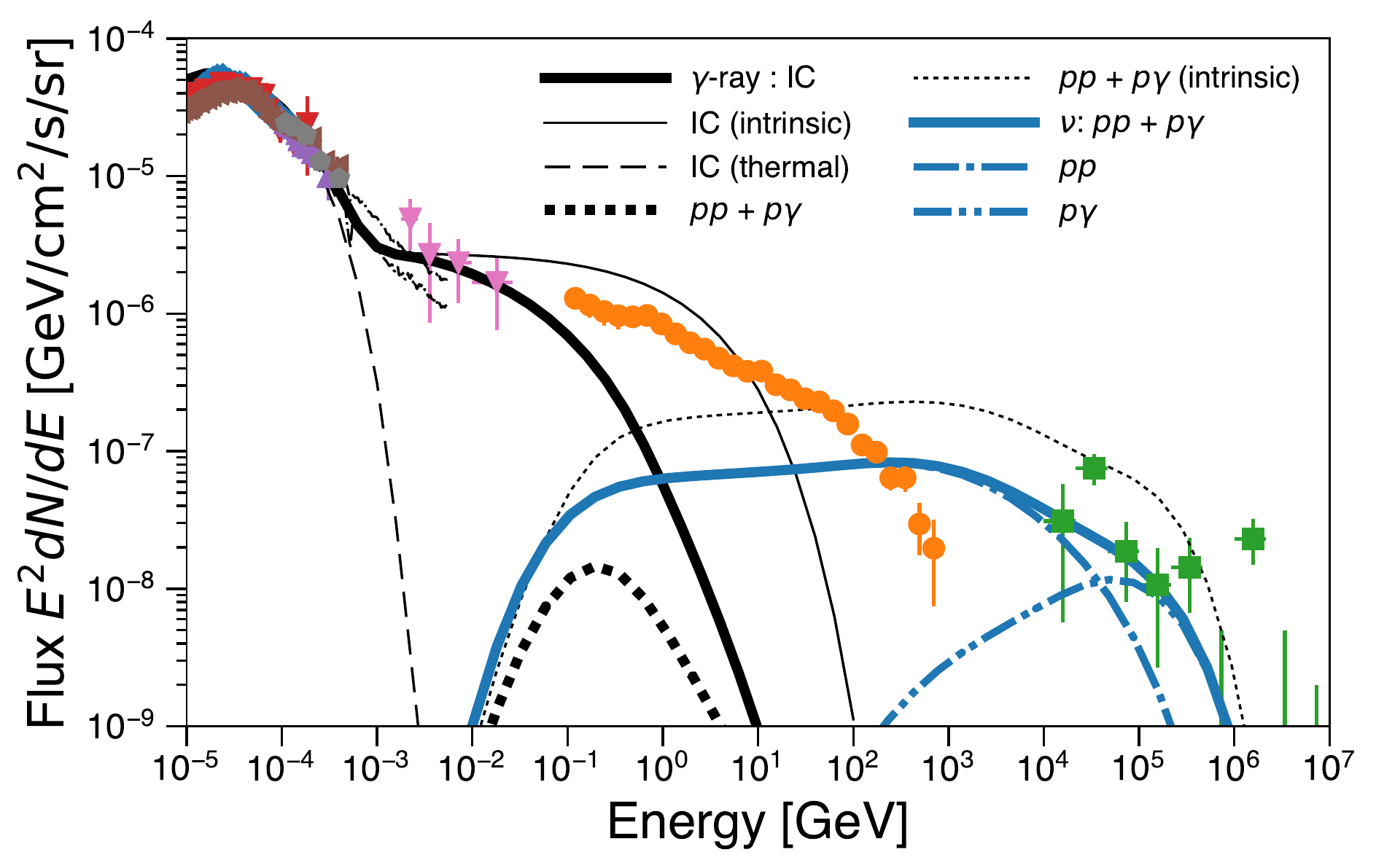}
\caption{The cosmic gamma-ray and neutrino background spectrum from \ac{agn} coronae with $p_{\rm inj}=2.0$ and $\eta_g=30$ assuming that the injection powers in electrons and protons are the same. The thick black solid and thick dot curves show the gamma-ray contribution of \ac{ic} interaction and \ac{pp}+\ac{pg} interaction, respectively, in which internal and EBL attenuation effects are taken into account. Corresponding thin curves show the spectra before the attenuation. The black dashed curve shows the \ac{ic} spectrum considering only thermal electrons. The blue dot-dashed, double-dot-dashed, and solid curve shows the neutrino contributions per flavour produced via \ac{pp} interactions, \ac{pg} interactions, and the sum of the two, respectively. The circle and square data points correspond to the total cosmic gamma-ray background spectrum measured by the {\it Fermi}/LAT \citep{Ackermann2015} and the cosmic neutrino background spectrum by the IceCube \citep{Aartsen2015}, respectively. The cosmic X-ray and MeV gamma-ray background spectrum data of  {\it HEAO}-1 A2 \citep{gru99}, {\it INTEGRAL} \citep{chu07}, {\it HEAO}-1 A4 \citep{kin97}, \textit{Swift}-BAT \citep{aje08}, {\it SMM} \citep{wat97}, Nagoya--Ballon \citep{fuk75}, COMPTEL \citep{wei00} are also shown in the figure. }\label{fig:CGNB_etag_100}
 \end{center}
\end{figure*}

\section{Cosmic Gamma-ray and Neutrino Background Fluxes From High Energy Particles in AGN Coronae}
\label{sec:background}

In this section, we calculate the cosmic gamma-ray and neutrino background spectra from AGN coronae. For the cosmological evolution of \acp{agn}, we follow \citet{Ueda2014} in which the evolutionary functions are defined at 2--10~keV intrinsic X-ray luminosity. We briefly review their formalism here. 

Based on the luminosity-dependent density evolution model, the \ac{agn} X-ray luminosity function at a given luminosity $L_X$ and a given redshift $z$ is defined as 
\begin{equation}
\frac{ d \Phi_{\rm X} (L_{\rm X}, z)}{ d{\rm log} L_{\rm X}} 
= \frac{ d \Phi_{\rm X} (L_{\rm X}, 0)}{ d{\rm log} L_{\rm X}} e(z, L_{\rm X}),
\end{equation}
where ${d \Phi_{\rm X} (L_{\rm X}, 0)}/{ d{\rm log} L_{\rm X}}$ is the luminosity function in the local universe defined as 
\begin{equation}
\frac{d \Phi_{\rm X} (L_{\rm X}, z=0)}{d{\rm log} L_{\rm X}} 
= A [(L_{\rm X}/L_{*})^{\gamma_1} + (L_{\rm X}/L_{*})^{\gamma_2}]^{-1},
\end{equation}
where $A$ is the normalization and $L_{*}$ is the break luminosity. $e(z, L_{\rm X})$ is the evolution factor represented as 
\begin{eqnarray}
  &&e(z, L_{\rm X} ) =  \\ \nonumber
&& \left\{ \begin{array}{ll}
     (1 + z)^{p1} & [z \le z_{c1}(L_{\rm X})], \\
     (1 + z_{c1})^{p1} 
     \left(\frac{ 1 + z}{ 1 + z_{c1}}\right)^{p2} & [z_{c1}(L_{\rm X}) < z \le z_{c2}], \\
     (1 + z_{c1})^{p1} \left(\frac{ 1 + z_{c2}}{ 1 + z_{c1}}\right)^{p2} \left(\frac{ 1 + z}{ 1 + z_{c2}}\right)^{p3} & [z > z_{c2}]. \\
  \end{array} \right.
\end{eqnarray}
Here the luminosity dependence for the $p1$ parameter is considered as 
\begin{equation}
p1(L_{\rm X})=p1^* + \beta_1 ({\rm log} L_{\rm X} - {\rm log} L_{\rm p}),
\end{equation}
where we set ${\rm log} L_{\rm p} = 44$. Both cutoff redshifts are given by power law functions of
$L_{\rm X}$ as
\begin{equation}
z_{\rm c1}(L_{\rm X}) = \left\{ \begin{array}{ll}
z_{\rm c1}^* (L_{\rm X}/L_{\rm a1})^{\alpha1} & [L_{\rm X} \le L_{\rm a1}], \\
z_{\rm c1}^* & [L_{\rm X} > L_{\rm a1}], \\
  \end{array} \right.
\end{equation}
and
\begin{equation}
z_{\rm c2}(L_{\rm X}) = \left\{ \begin{array}{ll}
z_{\rm c2}^* (L_{\rm X}/L_{\rm a2})^{\alpha2} & [L_{\rm X} \le L_{\rm a2}], \\
z_{\rm c2}^* & [L_{\rm X} > L_{\rm a2}]. \\
  \end{array} \right.
\end{equation}
The parameters are summarized in Table.~4 in \citet{Ueda2014}. There is also a substantial fraction of Compton-thick AGNs in the universe \citep[e.g.,][]{Ueda2003,Ricci2015}. In order to take into account this population, we multiply the normalization factor by a factor of 1.5 \citep[see][for details]{Ueda2014}.

The cosmic gamma-ray background fluxes are calculated as 
\begin{eqnarray}
\nonumber
    E^2\frac{dN}{dE} &=& \frac{c}{4\pi}\int\limits_{0.002}^5dz\int\limits_{41}^{47}d\log L_{\rm X} \left|\frac{dt}{dz}\right| \frac{ d \Phi_{\rm X} (L_{\rm X}, z)}{ d{\rm log} L_{\rm X}}\\ \nonumber
    &\times& \frac{L_\gamma(E', L_{\rm X})}{1+z}\frac{3u_{\rm int}(\tau_{\rm int}[E',\log L_{\rm X}])}{\tau_{\rm int}(E',\log L_{\rm X})}\\
    &\times&\exp(-\tau\EBL [E, z]),
\end{eqnarray}
where $E'=(1+z)E$ and $L_\gamma(E, L_{\rm X})$ is the gamma-ray luminosity at energy $E$ for a given X-ray luminosity of $L_{\rm X}$. The redshift and luminosity ranges are selected to be the same as in \citet{Ueda2014}. $\tau_{\rm int}$ and $\tau\EBL $ is the gamma-ray optical depth due to the internal photon field and the \ac{ebl}. We do not consider the cascade gamma-ray photons \citep[e.g.,][]{Inoue2012} because the gamma-ray energy fluxes due to hadronic interactions is already subdominant compairing to that by primary electrons. 

 The neutrino background fluxes can be also calculated in the same manner ignoring the gamma-ray attenuation terms and replacing $L_\gamma(E, L_{\rm X})$ with $L_\nu(E, L_{\rm X})$. $L_\nu(E, L_{\rm X})$ is the neutrino intensity at an energy of $E$ for a given X-ray luminosity of $L_{\rm X}$.

Figure~\ref{fig:CGNB_etag_100} shows the cosmic X-ray/gamma-ray and neutrino background spectra from \ac{agn} coronae assuming the case of $p_{\rm inj}=2.0$ and $\eta_g=30$. We also plot the observed background spectrum data by {\it HEAO}-1 A2 \citep{gru99}, {\it INTEGRAL} \citep{chu07}, {\it HEAO}-1 A4 \citep{kin97}, \textit{Swift}-BAT \citep{aje08}, {\it SMM} \citep{wat97}, Nagoya--Ballon \citep{fuk75}, COMPTEL \citep{wei00}, {\it Fermi}-LAT \citep{Ackermann2015}, and IceCube \citep{Aartsen2015}. 

\begin{figure}
 \begin{center}
  \includegraphics[width=9.0cm]{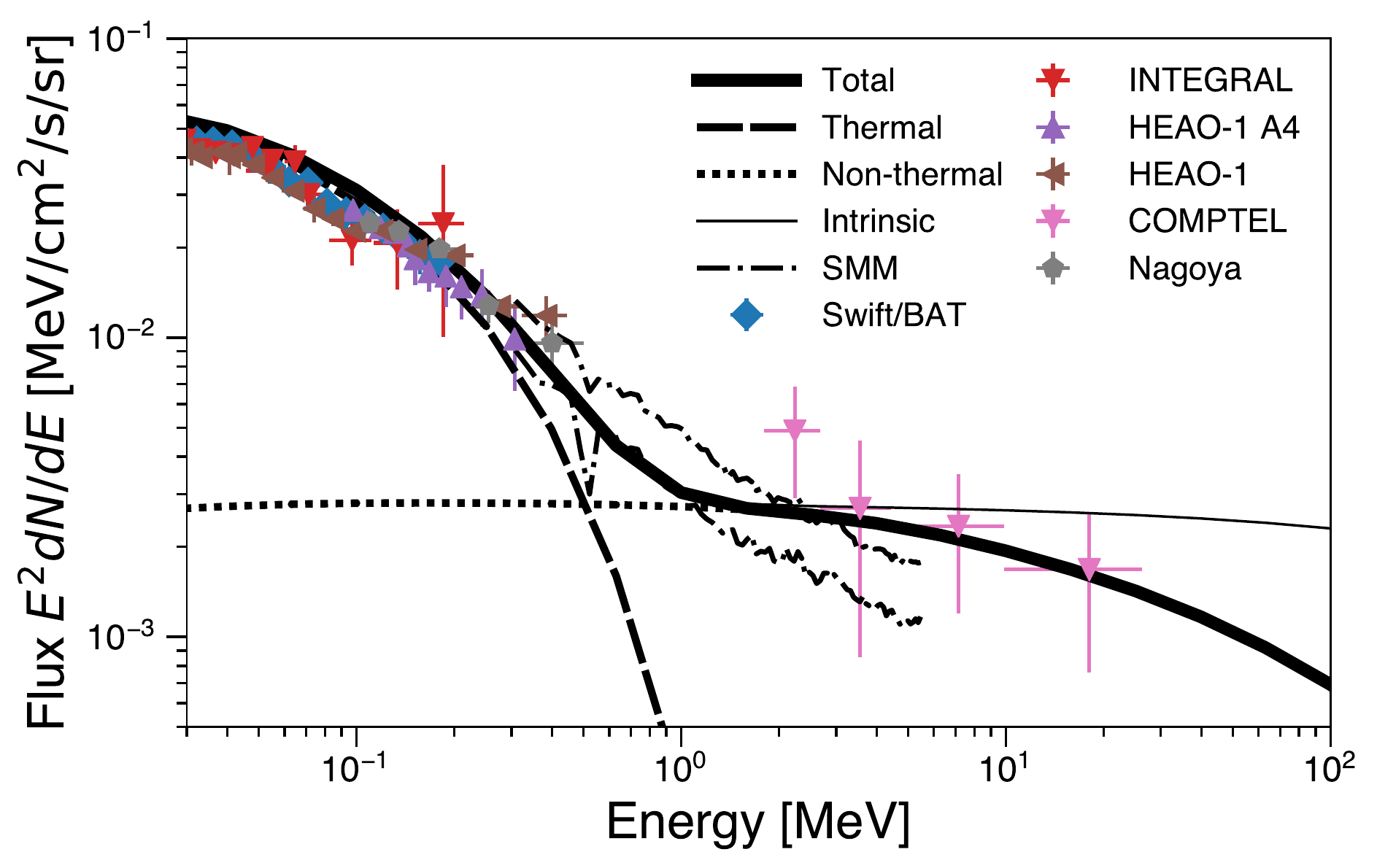}
\caption{Same as Figure~\ref{fig:CGNB_etag_100}, but enlarging the cosmic MeV gamma-ray background spectrum from 0.03~MeV to 100~MeV. The thick black solid  curve shows the total (thermal + non-thermal) contribution of \ac{ic} interaction where internal and EBL attenuation effects are taken into account. Thin curve shows the spectrum before the attenuation. The dashed and dotted curve shows the contribution from thermal electrons and non-thermal electrons, respectively. Contribution of reflection is included in the thermal contribution. The cosmic X-ray and MeV gamma-ray background spectrum data of  {\it HEAO}-1 A2 \citep{gru99}, {\it INTEGRAL} \citep{chu07}, {\it HEAO}-1 A4 \citep{kin97}, \textit{Swift}-BAT \citep{aje08}, {\it SMM} \citep{wat97}, Nagoya--Ballon \citep{fuk75}, COMPTEL \citep{wei00} are also shown in the figure.}\label{fig:CGB_MeV}
 \end{center}
\end{figure}

Figure~\ref{fig:CGB_MeV} shows the cosmic MeV gamma-ray background spectrum only from Figure~\ref{fig:CGNB_etag_100}. By setting $f_{\rm nth}=0.03$, the gamma-ray fluxes from \acp{agn} coronae due to \ac{ic} scattering by thermal and non-thermal electrons can nicely explain the observed cosmic MeV gamma-ray background radiation in an extension from the cosmic X-ray background radiation, which is known to be explained by Seyferts \citep{Ueda2014}. Since the spectral index of non-thermal electrons in the coronae is $\sim3$, the resulting MeV gamma-ray background spectrum becomes flat in $E^2dN/dE$ (See Fig.~\ref{fig:CGB_MeV}). Here, the cosmic X-ray background spectrum by Seyferts has a spectral cutoff above $\sim300$~keV because of temperature of thermal electrons $\sim100$~keV \citep{Ueda2014}. By summing up these two thermal and non-thermal components, superthermal tail appears in the sub-MeV band as observed by \citet[e.g.,][]{fuk75,kin97,wat97}. Since the dominant \ac{ic} contributors switches from thermal electrons to non-thermal electrons at around $1$~MeV, the MeV background spectrum may have spectral hardening feature at $\sim1$~MeV. In the figure, we set $\eta_g=30$. The result does not significantly change as far as $\eta_g<1000$. If $\eta_g>1000$, we may require lower $f_{\rm nth}$.

Due to the internal gamma-ray attenuation effect, these non-thermal gamma rays can not contribute to the emission above GeV. Because of the same reason, most of hadronic gamma-ray photons are attenuated by internal photon fields, resulting in generation of multiple secondary particles.
Since calculation of those populations are beyond the scope of this paper, we ignore those populations in our estimate. Moreover, as we describe above, the intrinsic hadronic fluxes are already an order of magnitude below the leptonic fluxes. Thus, pairs induced by hadronic cascades will not significantly change our results.

Here, \ac{ic} emission due to non-thermal electrons also contribute in the X-ray band. Their contribution is about $\sim5$\% at 30~keV of the observed cosmic X-ray background flux, which may reduce the required number of the Compton-thick population of AGNs. 

The model curve at $\sim10$~keV slightly overproduces the measured background spectrum. This is because we do not take into account X-ray attenuation by torus. However, the treatment of those soft X-ray photons does not affect our results at all.
 
\begin{figure}
 \begin{center}
  \includegraphics[width=9.0cm]{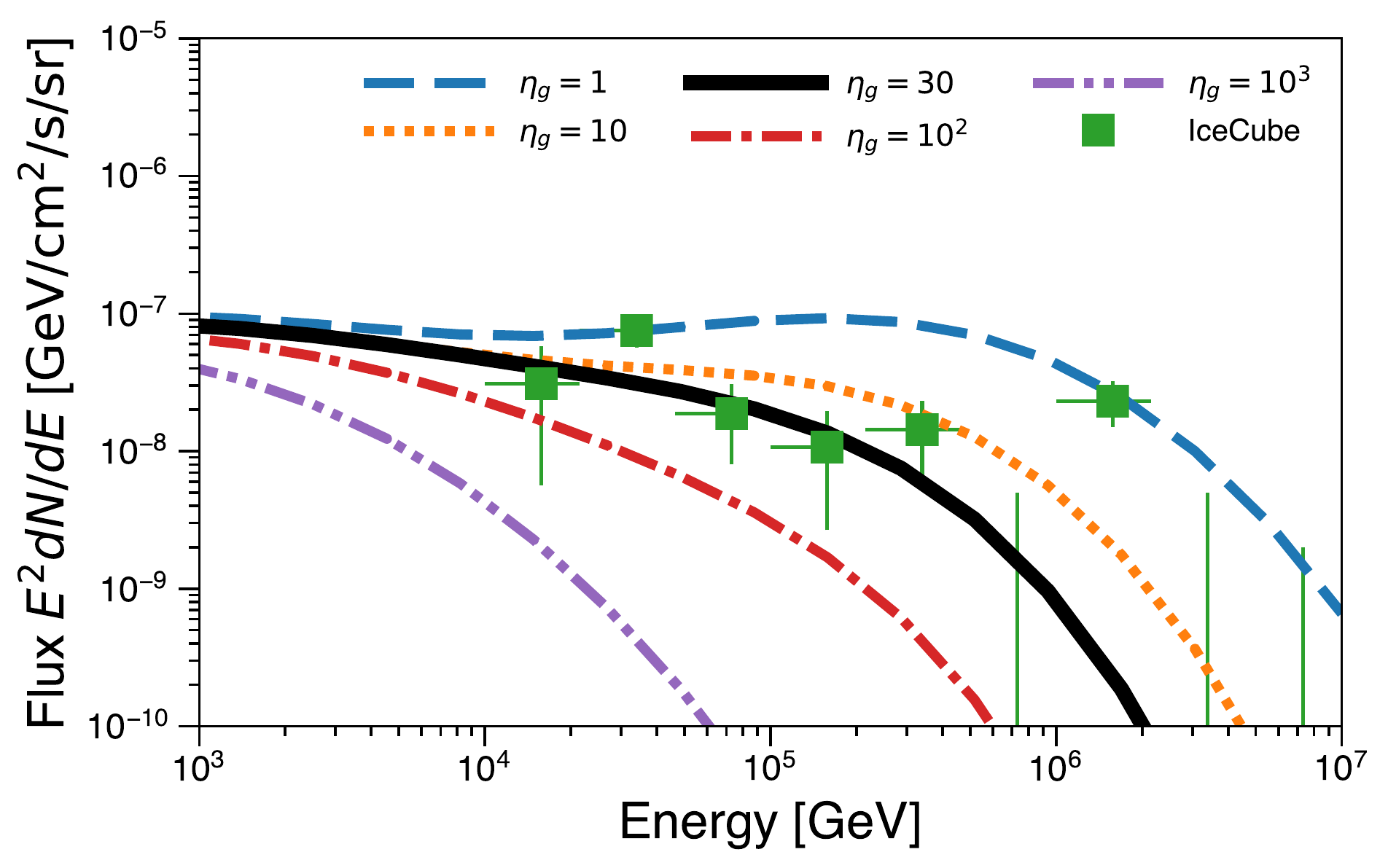}
\caption{The cosmic neutrino background spectrum per flavour from \ac{agn} coronae. The dashed, dotted, solid, dot-dashed, and double-dot-dashed curve shows the \ac{pp} + \ac{pg} contribution with $\eta_g=$1 (Bohm limit), 10, 30, $10^2$, and $10^3$, respectively. The square data points correspond to the cosmic neutrino background spectrum by the IceCube \citep{Aartsen2015}. }\label{fig:CNB_All}
 \end{center}
\end{figure}

For neutrinos, the combination of \ac{pp} and \ac{pg} interactions can nicely reproduce the IceCube fluxes below 100--300~TeV by assuming $\eta_g=30$ and about 5\% of the shock power into proton acceleration, same as electrons. \ac{pp} interactions dominate the flux at $\lesssim10$~TeV, while \ac{pg} interactions prevail above this energy. Because of the target photon field SED, \ac{pg} is subdominant in the GeV-TeV band. If we inject more powers into protons, it inevitably overproduces the IceCube background fluxes. As $\gtrsim$~GeV gamma rays are internally attenuated, AGN coronae emission will not be seen in GeV gamma-rays, even though they can make the IceCube neutrino fluxes. Such hidden cosmic-ray accelerators are suggested as a possible origin of the IceCube neutrinos \citep[see][for a general argument]{Murase2015}.

Figure~\ref{fig:CNB_All} shows the cosmic neutrino background spectra from \ac{agn} cores with various gyro factors ranging from 1 (Bohm limit) to $10^3$. It is clear that if $\eta_g\ll30$, the resulting neutrino fluxes overproduce the measured fluxes. On the contrary, if $\eta_g\gg30$, AGN coronae can not significantly contribute to the observed neutrino background fluxes. Thus, in order to explain the IceCube neutrino background fluxes by AGN cores, $\eta_g\sim30$ is required. However, we note that these estimates are based on the assumed energy injection fraction to protons. Recent particle-in-cell simulations of proton-electron plasma considering radiatively inefficient accretion flows (RIAFs) showed that protons will carry have several factors more energies than electrons \citep{Zhdankin2018}. If this is the case, larger $\eta_g$ is favored.

\section{Discussion}
\label{sec:discussion}

\subsection{Comparison with Previous works on High Energy Neutrinos}
In literature, it has been argued that high energy particles in the core of \acp{agn} generate intense neutrino emission \citep[e.g.,][]{Eichler1979,Begelman1990,Stecker1992,Alvarez-Muniz2004}. These originally predicted fluxes have been ruled out by high energy neutrino observations \citep{IceCube2005}. However, recent studies have revisited the estimated fluxes and found that \ac{agn} core models can account for the whole measured fluxes \citep{Stecker2013,Kalashev2015}. In this section, we would like to compare our results with those  recent studies \citep{Stecker2013,Kalashev2015}.  

The model suggest by \citet{Stecker2013} is very similar to the originally proposed one \citep{Stecker1992}, but  the background flux is assumed to be lower by a factor of 20. The original model is motivated by the models explaining \ac{agn} X-ray spectra by the electromagnetic cascade emission of secondary particles \citep{Zdziarski1986,Kazanas1986}, which is not the case based on current X-ray and gamma-ray observational results. The shock radius and the magnetic field strength was assumed to be $10R_s$ and $10^3$~G in the model by \citet{Stecker1992}. 

The model in \citet{Kalashev2015} is an extension of \citet{Stecker1992} taking into radial emission profile in the standard accretion disk for the consideration of the \ac{pg} cooling processes. In our modeling, we do not take into account such anisotropic radiation field. However, given the observationally determined corona size, the dominant photon targets are likely to be generated in the inner region of the coronae. The particle spectra in \citet{Kalashev2015} are fixed to match with the IceCube data. 

Neutrino fluxes or cosmic-ray spectra are fixed to match with the latest IceCube data in \citet{Stecker2013,Kalashev2015}. In this work, we take more physical approach. Corona plasma density, corona size, and magnetic field strength are determined from observations \citep{Inoue2018} in our work. For example, we set $R_c=40R_s$ and $B=10$~G based on ALMA observations \citep{Inoue2018}. With those parameters, we can follow the acceleration processes in coronae in the framework of \ac{dsa}. We found the AGN coronae can explain the IceCube neutrino background in the TeV band, if the gyrofactor is $\eta_g=30$ and about 5\% of the shock energy goes into proton acceleration. We also predict that next generation MeV gamma-ray and neutrino experiments can test our model by observing nearby bright Seyferts such as NGC~4151 and IC~4329A.

\subsection{Plasma Condition in Coronae}
Considering the plasma density in the accreting coronae, high energy particles may have sufficient time to redistribute their kinetic energy through thermalization by elastic Coulomb (EC) collisions before the gas reaches the event horizon \citep{Takahara1985,Mahadevan1997}. In this section, we discuss thermalization timescales of electrons and protons in the AGN coronae.

First, the electron thermalization timescale in the non-relativistic regime is estimated to be
\citep{Spitzer1962,Stepney1983}
\begin{eqnarray}
&&t_{\rm \ec, ee}\simeq \frac{4\sqrt{\pi}}{n_e\sigmat c\ln \Lambda} \theta_e^{3/2}\\ \nonumber
&&\simeq 1.1\times10^3\left(\frac{\taut}{1.1}\right)^{-1}\left(\frac{r_c}{40}\right)\left(\frac{\mbh}{10^8M_\odot}\right)\left(\frac{kT_e}{100\ {\rm keV}}\right)^{3/2}\ [{\rm s}],
\end{eqnarray}
where $\ln \Lambda\approx20$ is the Coulomb logarithm. For relativistic electrons with Lorentz factors $\gamma_e \gg 1 + \theta_e$ the thermalization timescale due to interactions with the background plasma becomes \citep{Dermer1989}
\begin{eqnarray}
\nonumber
t_{\rm \ec, ee}(\gamma_e)&=&\frac{4}{3} \frac{K_2(\theta_e^{-1})\gamma_e^3}{n_e\sigmat c(\ln \Lambda+9/16-\ln \sqrt{2})} \\
&\times&\left|\int\limits_1^\infty d\gamma_e'\exp(-u_{\rm ee})[\theta_e(1+2u)-\gamma_e]\right|^{-1},
\end{eqnarray}
where $K_n$ is the modified Bessel function of order $n$, and parameter \mbox{$u_{\rm ee}=(\gamma_e/\gamma'_e+\gamma'_e/\gamma_e)/2\theta_e$}. This equation can be approximated as 
\begin{eqnarray}
&& t_{\rm \ec, ee}(\gamma_e)\approx \\ \nonumber
&& \frac{2}{3} \frac{\gamma_e}{n_e\sigmat c(\ln \Lambda+9/16-\ln \sqrt{2})} \left|\frac{K_1(\theta_e^{-1})}{K_2(\theta_e^{-1})}-\frac{1}{\gamma_e} \right|^{-1}.
\end{eqnarray}
This is a good analytic approximation at $\theta_e\gtrsim0.3$ and $\gamma_e\gtrsim 2$  \citep{Dermer1989}.

Second, the proton-proton relaxation timescale in the non-relaticistic regime is estimated to be \citep{Spitzer1962,Stepney1983}
\begin{eqnarray}
\label{eq:t_ecpp}
&&t_{\rm \ec, pp}\simeq \frac{4\sqrt{\pi}}{n_p\sigmat c\ln \Lambda} \left(\frac{m_p}{m_e}\right)^{2}\theta_p^{3/2}\\ \nonumber
&&\simeq 4.7\times10^4\left(\frac{\taut}{1.1}\right)^{-1}\left(\frac{r_c}{40}\right)\left(\frac{\mbh}{10^8M_\odot}\right)\left(\frac{kT_p}{100\ {\rm keV}}\right)^{3/2}\ [{\rm s}],
\end{eqnarray}
where $\theta_p\equiv kT_p/m_pc^2$ is the dimensionless proton temperature. At high kinetic energies, nuclear interaction becomes important \citep[see][for details]{Gould1982}. In the mildly relativistic case, the elastic proton-proton relaxation timescale approximately becomes \citep{Gould1982}
\begin{equation}
t_{\rm \ec, pp}\simeq \frac{4}{n_p\sigma_hc}\frac{\beta_p\gamma_p^2}{\gamma_p^2-1},
\end{equation}
where $\sigma_h\sim2.3\times10^{-26}\ {\rm cm}^2$. This approximation is valid at $70\ {\rm MeV}\lesssim(\gamma-1)m_pc^2\lesssim500$~MeV. Above 500~MeV, inelastic processes start to dominate.

\begin{figure}
 \begin{center}
  \includegraphics[width=9.0cm]{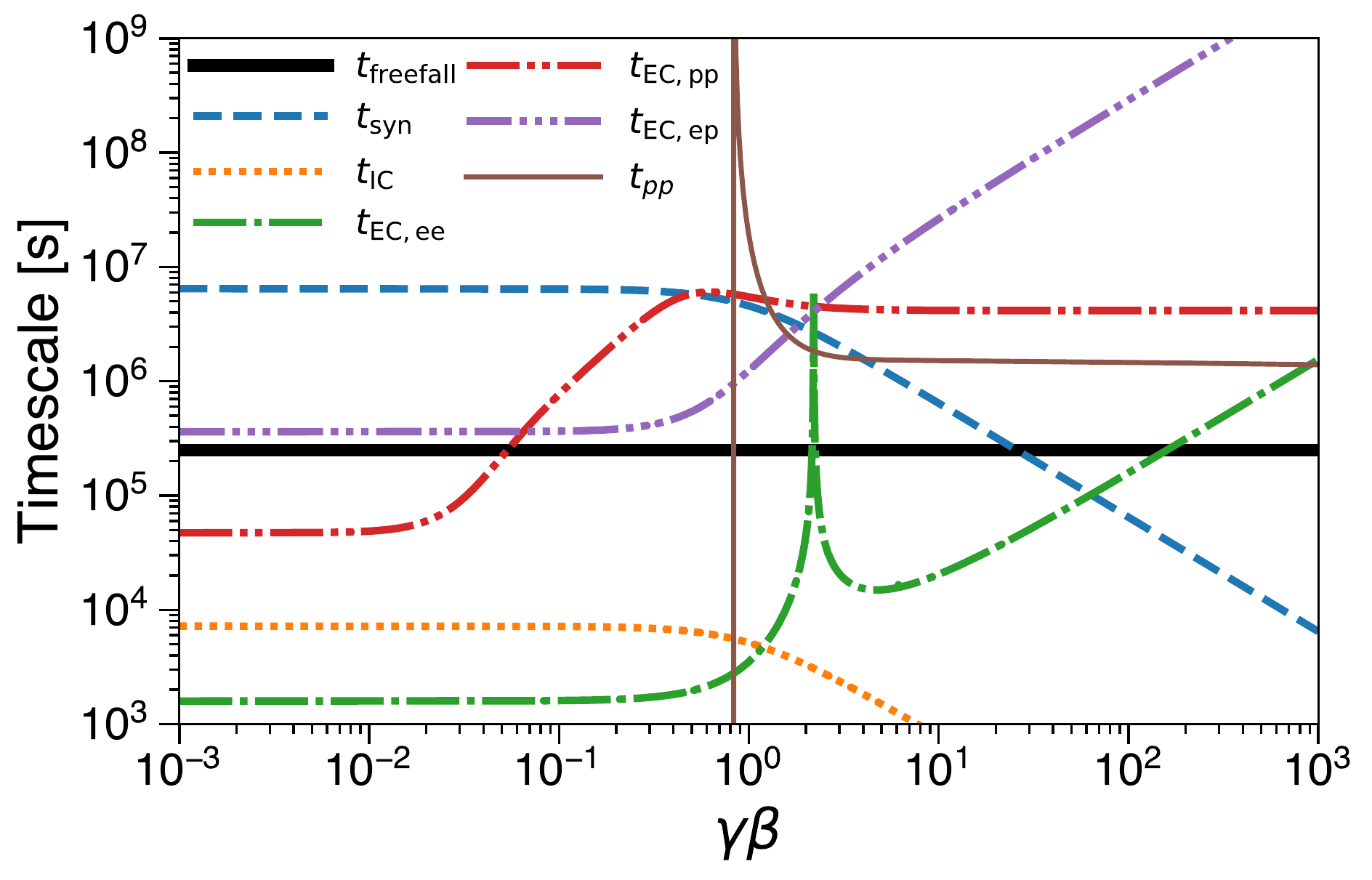}
\caption{Electron and proton thermalization timescales in AGN coronae together with radiative cooling and dynamical timescales. Thick solid curve shows the free-fall timescale. Dashed, dotted, and dot-dashed curve corresponds to synchrotron cooling, \ac{ic} cooling, and $ee$ \ac{ec} thermalization timescale for electrons, respectively. Double-dot-dashed, triple-dot-dashed, and thin solid curve corresponds to $pp$ \ac{ec} thermalization, $pe$ \ac{ec} thermalization, and $pp$ inelastic interaction timescale for protons, respectively. We set $\log L_X=44$, $\taut=1.1$, $R_c=40R_s$, and $kT_e=kT_p=100$~keV.}\label{fig:time_ec}
 \end{center}
\end{figure}

Lastly, the proton-electron thermalization timescale due to \ac{ec} collisions in the non-relativistic regime is estimated to be \citep{Spitzer1962,Stepney1983}
\begin{eqnarray}
&&t_{\rm \ec, ep}\simeq \frac{\sqrt{\pi/2}}{n_e\sigmat c\ln \Lambda} \left(\frac{m_p}{m_e}\right)(\theta_e+\theta_p)^{3/2}\\ \nonumber
&&\gtrsim 3.6\times10^5\left(\frac{\taut}{1.1}\right)^{-1}\left(\frac{r_c}{40}\right)\left(\frac{\mbh}{10^8M_\odot}\right)\left(\frac{kT_e}{100\ {\rm keV}}\right)^{3/2}\ [{\rm s}],
\end{eqnarray}
where we assume $\theta_p=\theta_e$. The temperature of a hot accretion can roughly reach to virial temperature $T_p\simeq G\mbh m_p/3kR\sim3\times10^{12}(R/R_s)^{-1}$~K. At such higher temperature, $t_{\rm \ec, ep}$ becomes longer. In the case of relativistic protons, the energy loss timescale through \ac{ec} interactions is given as \citep{Mannheim1994,Dermer1996}
\begin{equation}
t_{\rm \ec, ep}\simeq 1.2\times10^3 \frac{(3.8\theta_e^{3/2}+\beta_p^3)(\gamma_p-1)}{n_p\sigmat c\beta_p^2\ln\Lambda},
\end{equation}
where $\beta_p = \sqrt{1-1/\gamma_p^2}$. At $\gamma_p\gg1$ and $\theta_e\ll1$, the relativistic \ac{ec} scattering relaxation time can be approximated as 
\begin{equation}
t_{\rm \ec, ep}\simeq 2.9\times10^8 \left(\frac{\taut}{1.1}\right)^{-1}\left(\frac{r_c}{40}\right)\left(\frac{\mbh}{10^8M_\odot}\right) \left(\frac{\gamma_p}{100}\right)\ [{\rm s}].
\end{equation}

Fig.~\ref{fig:time_ec} shows EC thermalization timescales for electrons and protons for the luminosity of $L_X=10^{44}~{\rm erg\ s^{-1}}$. Since EC thermalization is effective at low energy particles, the horizontal axis is shown in $\gamma\beta$. 

Around $\gamma_e\beta_e\sim2$, $t_{\ec,ee}$ shows a sharp feature, which is related to the temperature of the background plasma, $kT_e=100$~keV. At this temperature, the electron distribution has a peak around $\sim3kT_e$ corresponding to $\gamma_e\beta_e\sim1.2$. Thus, around this energy, mean energy transfer is small. We note that below this energy, electrons gain energies from the background plasma through elastic $ee$ scatterings rather than loosing their energies \citep{Dermer1989}, however, this energy gain process is not considered in our work, since it is not relevant for our energy range of interest. As seen in the Fig.~\ref{fig:time_ec}, the energy loss process of electrons is dominated by the Compton cooling at $\gamma_e\beta_e\gtrsim1$. 

Following \citet{Gould1982}, we calculate the elastic $pp$ timescale in the mildly relativistic regime. Since it assumes an incident proton has much higher kinetic energy than background plasma, we combine the non-relativistic $t_{\ec,pp}$ (Equation. \ref{eq:t_ecpp}) and that from \citet{Gould1982}. As discussed above, inelastic processes start to dominate at the kinetic energies of $\gtrsim500$~MeV ($\gamma_p\beta_p\gtrsim1.2$). For the comparison, we also show inelastic $pp$ interaction timescale $t_{pp}$. 

As the proton-electron Coulomb timescale ($t_{\ec,pe}$) is longer than $t_{\rm fall}$, protons and electrons may not be in the thermal equilibrium in AGN coronae. The proton temperature of a hot accretion can roughly reach to virial temperature $T_p\simeq G\mbh m_p/3kR\sim3\times10^{12}(R/R_s)^{-1}$~K, which is $\gg T_e$. And, the existence of pairs in coronae can reduce $n_p$. Moreover, the shock heated proton temperature becomes $kT_{p}\sim 3m_p v_{\rm sh}^2\sim 4(r_c/40)^{-1}~{\rm MeV}$. Those shock heated protons and electrons also gain and loose their energies through the processes and would contribute as a thermal population in the coronae. These electrons are heated and cooled through EC proton-electron thermalization and Comptonization, respectively \citep[e.g.,][]{Katz2011,Murase2011}. The heating rate can be written as
\begin{equation}
-\frac{dT_p}{dt}=\frac{dT_e}{dt}=\frac{T_p}{t_{\ec,pe}}\simeq \frac{n_e\sigmat c\ln \Lambda}{\sqrt{\pi/2}} \left(\frac{m_e}{m_p}\right)T_p\theta_e^{-3/2},
\end{equation}
assuming $\theta_e \gg \theta_p$. The cooling rate through Comptonization is
\begin{equation}
\frac{dT_e}{dt}\approx-\frac{4}{3}	\frac{\sigma_T U_{\rm ph,tot}T_e}{m_ec}
\end{equation}
By equating these two heating and cooling rates of thermal electrons, the shock heating electron temperature is estimated to be 
\begin{eqnarray}
kT_e&\simeq& k\left(\frac{3\ln \Lambda}{4\sqrt{\pi/2}}\frac{m_e}{m_p}\frac{n_e}{U_{\rm ph,tot}}T_p\right)^{2/5}\\ \nonumber
&\simeq&86\left(\frac{\taut}{1.1}\right)^{2/5}~[\rm keV],
\end{eqnarray}
where we assume $L_{\rm ph, bol}\propto M_{\rm BH}$. This temperature is close to the measured coronal temperature. Therefore, such shock heating mechanism may be able to explain the current observed coronal temperature. For the understanding the detailed nature of thermal coronae, further studies including thermodynamical processes are required.

\subsection{Other Particle Acceleration Mechanisms}
\label{sec:other_acc}
In this paper, we consider the \ac{dsa} as fiducial acceleration mechanism. However, other acceleration mechanisms such as turbulent acceleration, magnetosphere acceleration, and magnetic reconnection can also operate in AGN coronae. We briefly discuss these processes here.

First, turbulent acceleration is considered for low-accretion rate objects such as low-luminosity \acp{agn} \citep[e.g.,][]{Kimura2015,Zhdankin2017,Zhdankin2018,Wong2019}. In this scenario, particles are accelerated stocastically by turbulence and magnetic reconnection in accretion disk or coronae. Recently, \citet{Zhdankin2018} investigated electron-ion plasma energization via turbulent dissipation in RIAFs using particle-in-cell simulations for the ion temperature $T_i$ in the range of $m_ec^2\lesssim k_BT_i\lesssim m_pc^2$. Turbulent electron-ion plasma driven by MRIs generate power-law spectra for both species and the indices depends on the initial ion temperature. The fraction of the kinetic energy in the non-thermal ions and electrons are $\sim60$\% and $6$\% for ions and electrons at $k_BT_i\sim m_ec^2$, respectively. The fraction in non-thermal electrons is close to the required value for the MeV background (See \S~\ref{sec:background}).

We briefly follow the stochastic acceleration in the AGN coronae case. According to the quasi-linear theory, the diffusion coefficient in the momentum space is \cite[e.g.,][]{Dermer1996}
\begin{equation}
D_p\simeq (m_pc)^2 (ck_{\rm min})\left(\frac{\vA }{c}\right)^2\zeta(r_\lar k_{\rm min})^{q-2}\gamma^q,
\end{equation}
where $k_{\rm min}\sim R_c^{-1}$ is the minimum wave number of turbulence spectrum (corresponding to the size of the corona), $\vA =B/\sqrt{4\pi m_p n_p}$ is the Alfv\'en speed, $r_\lar=m_pc^2/eB$ is the Larmor radius, and $\zeta=\delta B^2/B^2$ is the ratio of strength of turbulence fields against the background. Then, the acceleration timescale is estimated to be
\begin{equation}
t_{\STO}\simeq\frac{p^2}{D_p}\simeq\frac{1}{\zeta}\left(\frac{\vA }{c}\right)^{-2}\frac{R_c}{c}\left(\frac{r_\lar}{R_c}\right)^{2-q}\gamma^{2-q}
\end{equation}

Assuming the Kolomogorov spectrum for the turbulent ($q=5/3$) and $\zeta=1$, the timescale becomes
\begin{eqnarray}
\nonumber
t_{\STO}&\simeq& 3.1\times10^7\left(\frac{\taut}{1.1}\right)\left(\frac{r_c}{40}\right)^{-1/3}\left(\frac{\mbh}{10^8M_\odot}\right)^{-1/3}\\ 
&\times&\left(\frac{B}{10\ {\rm G}}\right)^{-7/3}\left(\frac{\gamma_p}{100}\right)^{1/3}\ [{\rm s}].
\end{eqnarray}
Thus, stochastic acceleration appears to be inefficient as compared to the typical cooling rates. This is caused by the measured weak magnetic fields, which results in small Alfv\'en speed. If the magnetic fields are amplified by MRIs, more efficient acceleration can be realized \citep[e.g.,][]{Zhdankin2018}\footnote{After we submitted our paper to the journal and arXiv, similar study on AGN coronae by \citet{Murase2019} appeared on  arXiv. Both studies are independent and the most different point is the assumed particle acceleration processes. In our paper, we consider \ac{dsa}, while \citet{Murase2019} consider stochastic acceleration motivated by recent numerical simulations \citep{Kimura2019}. However, as we discussed in this section, stochastic acceleration may not work given the ALMA results of weak coronal magnetic field.}.

Second, magnetosphere acceleration can also accelerate particles in the vicinity of \acp{smbh} \citep[e.g.,][]{Beskin1992,Levinson2000,Neronov2007,Levinson2011,Rieger2011}. At low accretion rates, the injection of charges into the BH magnetosphere is not sufficient for a full screening of the electric field induced by the rotation of the compact object. The regions with unscreened electric field, so-called gaps, are able to accelerate charged particles effectively. 

In order to have gaps, the maximum allowed accretion rate is \citep{Levinson2011,Aleksic2014,Aharonian2017}
\begin{equation}
\dot{m}<3\times10^{-4}\left(\frac{\mbh}{10^8M_\odot}\right)^{-1/7},
\end{equation}
where $\dot{m}$ is the accretion rate in the Eddington units. Since we are considering the standard accretion disk regime $\dot{m}\gtrsim0.01$, particle acceleration by gaps will not be operated in our case.

Lastly, magnetic reconnection would accelerate particles \citep[see e.g.,][for reviews]{Hoshino2012}. Reconnection would naturally happens in coronae as they are magnetized and  radiative magnetic reconnection is suggested as an origin of the X-ray emission seen in accreting black hole systems \citep{Beloborodov2017}. However, even in the case of solar flares, particle acceleration mechanisms in magnetic reconnection is still uncertain \citep[e.g.,][]{Liu2008,Nishizuka2013}. Although quantitative discussion is not easy here, the available energy injection power can estimated as
\begin{eqnarray}
	P_B &=& \frac{B^2R_c^2v_A}{2}\\ \nonumber
	&\simeq &5.4\times10^{39}\left(\frac{\taut}{1.1}\right)^{-1/2}\left(\frac{r_c}{40}\right)^{5/2}\left(\frac{\mbh}{10^8M_\odot}\right)^{5/2}\\ \nonumber
	&\times&\left(\frac{B}{10~{\rm G}}\right)^3\ [{\rm erg\ s^{-1}}].
\end{eqnarray}
This power is not sufficient for providing the non-thermal particle energies. For detailed estimation, we may need to consider spatial distribution fo magnetic field. However, such information is not currently available.

\subsection{Cosmic MeV Gamma-ray Background Radiation}
It is known that Seyferts generate the cosmic X-ray background radiation \citep{Ueda2014}. The cosmic gamma-ray background at 0.1--820~GeV is believed to be explained by three components: blazars \citep[e.g.,][]{Inoue2009,Ajello2015}, radio galaxies \citep{Inoue2011}, and star-forming galaxies \citep{Ackermann2012_SB}, even though the contributions of radio galaxies and star-forming galaxies are still uncertain due to a small number of gamma-ray detected samples. On the contrary to the cosmic X-ray and GeV background radiation, the origin of the cosmic MeV gamma-ray background radiation is still veiled in mystery. 

As a possible scenario, non-thermal \ac{ic} emission from coronae in Seyferts has been suggested \citep{Inoue2008}. The MeV tail extended from the X-ray background spectrum is generated by non-thermal electrons with very soft spectral index \citep{Inoue2008}. However, non-thermal electrons are included in an ad hoc way. In our work, we consider the particle acceleration and cooling processes given the latest observations. The tail is due to the superposition of thermal Comptonization cut-off spectrum and $\gamma\gamma$ attenuated flat non-thermal \ac{ic} component. We can distinguish these two scenarios by observing individual objects in radio and X-ray bands.

Not only Seyferts, but also blazars are considered as a candidate as the origin of the MeV background \citep{Ajello2009}. In order to distinguish Seyferts and blazars, we need to resolve the MeV sky. However, it is not easy even with future MeV instruments \citep{Inoue2015}. Here, it is suggested that anisotropy measurements may distinguish these two scenarios \citep{Inoue2013_CXB} because blazar background should feature stronger Poisson fluctuations. Future MeV gamma-ray anisotropy observations will be important to understand the particle acceleration in coronae and the origin of the MeV gamma-ray background radiation.

\subsection{Gamma-ray Observations toward Seyferts}
Gamma rays from Seyfert galaxies are not robustly detected yet \citep{Lin1993,Teng2011,Ackermann2012}. Possible signature of gamma-ray emission above 0.1~GeV have been reported for ESO~323-G077 and NGC~6814 \citep{Ackermann2012}, whose X-ray luminosities are about $10^{43}\ {\rm erg\ s^{-1}}$. The required luminosity ratio between X-ray and gamma-ray $L_{\rm 0.1-10~GeV}/L_{\rm 14-195~keV}$ for these sources is about 0.1 \citep{Ackermann2012}. Our model estimates this ratio as $\sim0.01$. Therefore, coronal gamma-ray emission is most-like not able to account for the observed gamma-ray fluxes from those Seyfert galaxies.

Although gamma rays from other Seyferts have not been detected yet, {\it Fermi}/LAT has set upper limits on their gamma-ray fluxes \citep{Teng2011,Ackermann2012}. Based on the analysis of the first 2-3 years data, $L_{\rm 0.1-10~GeV}/L_{\rm 14-195~keV}<0.1$ in the 95\% confidence level is obtained in most cases, which is consistent with our model estimate. The most stringent observational constraint is derived for NGC~4151, in which $L_{\rm 0.1-10~GeV}/L_{\rm 14-195~keV}<0.0025$, even though the limit can vary with an assumed spectral shape. Following our models, the current 10~yrs survey data of {\it Fermi}/LAT may be able to see NGC~4151 (Figure.~\ref{fig:AGN_SED}), even though the expected flux is almost at the sensitivity limit.

\subsection{Fraction of Non-thermal Electrons}
We set the energy fraction of non-thermal electrons in AGN coronae as $f_{\rm nth}=0.03$ because it nicely reproduces the observed MeV gamma-ray background radiation. As discussed in \citet{Inoue2018}, $f_{\rm nth}$, $B$, and $R_c$ are closely tied, current radio and X-ray data do not allow us to solve these three parameters simultaneously without decoupling thermal and non-thermal components. 

Observationally, $f_{\rm nth}$ is constrained as $<0.3$ in order not to violate X-ray data based on {\it NuSTAR} observations \citep{Fabian2017}. If $f_{\rm nth}$ is significantly lower, it becomes difficult for Seyfert to explain the MeV gamma-ray background radiation. However, too much lower $f_{\rm nth}$ contradicts with other observations since it requires a bigger $R_c$ based on the radio spectral fitting. If we set $f_{\rm nth}=10^{-3}$ and $10^{-4}$, $R_c$ becomes $\sim70R_s$ and $\sim100R_s$, respectively. The size of coronae is also constrained as an order of $\sim10R_s$ by optical--X-ray spectral fitting studies \citep{Jin2012} and micorolensing observation \citep{Morgan2012}. Therefore, $f_{\rm nth}$ can not become much smaller than the adopted value. 

\subsection{Nuclear Spallation in AGNs}
Given the \ac{alma} results, particle accelerations occurs in AGN coronae. As we demonstrated, high energy protons are easily accelerated in coronae. These high energy protons can be also traced by future high-resolution calorimeter spectroscopy in the X-ray band such as {\it XRISM} \citep{Tashiro2018} and {\it Athena} \citep{Nandra2013}\footnote{The Athena X-ray observatory website (\url{https://www.the-athena-x-ray-observatory.eu/}}. As narrow line features are seen in AGN X-ray disk spectra, there are abundant metal elements in AGN cores. Accelerated protons also interact with those nuclei and induce nuclear spallation. The nuclear spallation in AGN disks will result in enhancement of emission lines from Mn, Cr, V, and Ti \citep{Gallo2019}. Those signatures will be another clue for the test of our model.

\section{Conclusion}
\label{sec:conclusion}
Recently, \citet{Inoue2018} has reported the coronae of Seyferts are composed of both thermal and non-thermal electrons based on \ac{alma} observations, which implies that particle acceleration occurs in AGN coronae. In order to investigate the production mechanism of those high energy particles, we study the particle acceleration process in AGN coronae. We consider particle acceleration by the \ac{dsa} process in the coronae as an example. By taking into account the observationally determined coronal properties, such as temperature, density, size, and magnetic field strength, we found that standard \ac{dsa} processes can easily reproduce the observed non-thermal electron in the coronae with an injection electron spectral index of $p_{\rm inj}=2$. Even in low acceleration efficiency cases ($\eta_g\sim10^6$), such populations can be realized in coronae. Given the observed magnetic field strength of 10~G and accretion rates, we also found that other possible acceleration mechanisms such as turbulent acceleration, magnetosphere acceleration, and magnetic reconnection confront difficulty in reproducing the observed non-thermal electrons. 

The accelerated non-thermal electron populations will generate a MeV gamma-ray power-law spectrum in the AGN SEDs up to $\sim0.1$~GeV, which is limited by internal gamma-ray attenuation. In the sub-MeV band, the spectrum shows a superthermal tail due to the combination of thermal and non-thermal components and spectral flattening occurs at $\sim1$~MeV. These superthermal and flat spectral tails should be tested by future MeV gamma-ray missions.

We also study the contribution of \ac{agn} coronae to the cosmic gamma-ray background radiation. By setting the energy fraction of non-thermal electrons $f_{\rm nth}\sim3$\%, corresponding to $\sim5$\% of the shock energy in electron acceleration, \ac{agn} coronae can explain the MeV background in an extension of the X-ray background contribution of Seyferts. Due to a strong internal gamma-ray attenuation effect, the contribution of \ac{agn} coronae to the GeV background is negligible. 

Accelerated particles would also result in neutrino production through hadronic processes. Intense neutrino emission has been expected to be produced in \ac{agn} coronae once hadrons are accelerated together \citep[e.g.,][]{Begelman1990,Stecker1992,Alvarez-Muniz2004}. Recent studies have proposed that these \ac{agn} core models could reproduce the high energy neutrino fluxed measured by IceCube \citep{Stecker2005,Stecker2013,Kalashev2015}. However, normalization of neutrino fluxes from \acp{agn} and acceleration properties of high energy particles in those models are assumed to match with the observation. 

We found that \ac{agn} coronae can explain the diffuse neutrino fluxes below 100--300~TeV under specific parameters of energy injection rates in protons and gyro factors. The allowed parameter regions are quite narrow. Protons and electrons should have the same energy injection rate and the gyro factor $\eta_g$ should be $\sim30$. IceCube Gen-2 will be able to test this scenario by searching the neutrino signal from nearby Seyfert galaxies such as NGC~4151 and IC~4329A. 

In summary, Seyfert coronae are feasible sites for particle acceleration. If the energy injection rate is 5\% for both protons and electrons and the gyro factor is $\eta_g=30$, they may be able to simultaneously explain the cosmic X-ray, MeV gamma-ray, and TeV neutrino background radiation. Future MeV gamma-ray and TeV neutrino observations will be able to test this scenario by observations of nearby bright Seyferts.

\acknowledgments
We thank the anonymous referee for his/her helpful comments which improved the manuscript. We also would like to thank Tsuguo Aramaki, Mitch Begelman, Norita Kawanaka, Shigeo Kimura, Ari Laor, Kohta Murase, Satomi Nakahara, and Marek Sikora for useful discussions and comments. YI is supported by JSPS KAKENHI Grant Number JP16K13813, JP19K14772, program of Leading Initiative for Excellent Young Researchers, MEXT, Japan, and RIKEN iTHEMS Program. DK is supported by JSPS KAKENHI Grant Numbers JP18H03722, JP24105007, and JP16H02170. 

\bibliography{references}

\listofchanges

\end{document}